\documentclass[useAMS,usenatbib,onecolumn]{mn2e}
\usepackage{epsf}
\usepackage{graphicx}
\usepackage{epstopdf}
\usepackage{url}
\usepackage{hyperref}
\title[Faraday Conversion In Relativistic Plasmas]
{Faraday Conversion And Rotation In Uniformly Magnetized Relativistic Plasmas}   
\author[Huang and Shcherbakov]
{Lei Huang$^{1,2}$\thanks{E-mail: muduri@shao.ac.cn} and Roman V. Shcherbakov$^{3}$\\
$^{1}$Key Laboratory for Research in Galaxies and Cosmology, Shanghai Astronomical Observatory, Chinese Academy of Sciences, \\
 \quad Shanghai 200030, China\\
$^{2}$Key Laboratory of Radio Astronomy, Chinese Academy of Sciences\\
$^{3}$Harvard-Smithsonian Center for Astrophysics, 60 Garden Street, Cambridge, MA 02138, USA}

\pagerange{\pageref{firstpage}--\pageref{lastpage}} \pubyear{2011}

\begin{document}
\maketitle
\label{firstpage}

\begin{abstract}
We provide precise fitting formulae for Faraday conversion and rotation coefficients in uniformly magnetized relativistic plasma.  The formulae are immediately applicable to Rotation Measure and Circular Polarization (CP) production in jets and hot accretion flows.   
We show the recipe and results for arbitrary isotropic particle distributions, in particular thermal and power-law. 
The exact Faraday conversion coefficient is found to approach zero with the increasing particle energy. The non-linear corrections of Faraday conversion and rotation coefficients are found essential for reliable CP interpretation of Sgr A*.  
\end{abstract}


\section{INTRODUCTION}
The cyclo-synchrotron emission, also called magneto-bremsstrahlung emission, is one of the most important radiative mechanism in astrophysics. It is believed to produce radio emission in the centers of AGNs and LLAGNs (low luminosity AGNs, such as the Galactic Center).  A polarized nature of cyclo-synchrotron emission is of increasing interest for radio observers. With the help of polarization one can understand the magnetic field structure in radio sources. Basic theory of emission and propagation of polarized light has been established \citep[e.g.,][]{legg, pach, jones}. Particles in cold plasma emit cyclotron radiation, which is circularly polarized (CP). When linearly polarized (LP) light propagates through cold magnetized plasma, it undergoes Faraday rotation. In turn, relativistic plasma emits synchrotron radiation, which is linearly polarized. Light traversing relativistic plasma undergoes both Faraday conversion and Faraday rotation.

In simple theory the strength of Faraday rotation effect is proportional to $\lambda^2 n_e {\bmath B} \cdot \delta {\bmath l}$, where $\lambda$ is the photon wavelength, $n_e$ is electron density, ${\bmath B}$ is magnetic field vector, and $\bmath l$ is the displacement along the line of sight. However, \citet{trubnikov} and \citet{melrose97a} have shown that in a general case Faraday rotation of plasma depends also on the Lorentz factor of electrons $\gamma$. Faraday rotation weakens with the increase of $\gamma$ as $\ln \gamma /\gamma^2$.  The electric vector position angle (EVPA) of LP light will be preserved better, if the electrons are relativistic. Thus we can infer the intrinsic EVPA of a synchrotron-emitting source. The electrons in the vicinity of a black hole Sagittarius A* (Sgr A*) in Galactic Center are often modeled by a relativistic Maxwellian (thermal) distribution with temperature $> 10^{10}$K \citep[e.g.][]{yuanf03}. When Faraday rotation is strong and non-uniform across the beam, then beam depolarization can diminish the resultant LP fraction \citep{bower05}. The effect of large Lorentz factors on Faraday rotation measure near Sgr A* must be considered to explain the detected LP fluxes in sub-millimeter to near-infrared bands \citep{qua_gru}.    Previous work \citep[e.g.][]{ginzb, sazonov, melrose97b} also highlighted Faraday conversion, or generalized Faraday rotation. This quantity describes the interconversion of linearly and circularly polarized light.  It is normally expected \citep{homan} to convert emitted LP light into CP light during propagation in magnetized medium. This is the likely cause of observed high CP fraction of Sgr A* spectrum in radio band \citep{bower02} and sub-millimeter band \citep{munoz}. 

The strength of Faraday conversion was found proportional to $\lambda^3 n_e B^2 \sin^2\theta \delta l$, where $\theta$ is the angle between $\bmath B$ and $\delta {\bmath l}$ (or $\bmath k$, see in Figure \ref{figvector}). Additional suggested proportionality to electron temperature $T_e$ makes Faraday conversion reach very large values in relativistic plasmas. However, this proportionality ceases at very high $T_e$ and Faraday conversion measure approaches zero \citep{shcher_farad}. A detectable CP fraction can be generated near Sgr A* in sub-millimeter band \citep[e.g.][]{ballantyne}, but precise treatment of Faraday rotation and conversion is essential \citep{shcher_appl}. A non-zero CP fraction is already detected with SMA (Munoz et al., 2011, ApJ, submitted) at $230$~GHz and $345$~GHz. This by itself points in the direction of a very hot radiatively inefficient accretion flow (RIAF). The origin of CP near Sgr A* and jets was quantified by \citet{beckert03, shcher_appl}. Yet more detailed and accurate calculations are needed to quantify the circular polarizations for the Galactic Center (GC) supermassive black hole and other radio source.

\citet{huang09a} was the first to incorporate Faraday rotation and conversion within general relativistic (GR) polarized radiative transfer framework, though with some approximations. \citet[][hereafter Paper I]{shcher_huang} provided a method to accurately calculate cyclo-synchrotron absorption and Faraday conversion/rotation for electrons in isotropic thermal distribution and outlined the exact procedure for polarized radiative transfer in GR. Precise Faraday rotation and conversion coefficients were computed earlier for thermal plasmas in \citet{shcher_farad}.  An important cornerstone in computing propagation effects is linear approximation, in which only the first non-zero terms in series expansion in $\Omega_0/\omega$ ratio are taken for correspondent quantities (e.g. formula~\ref{lin_approx}). It was also derived therein and found consistent with result provided by \citet{melrose97b}. The precise values of Faraday conversion coefficient in \citet{shcher_farad} and Paper I match the linear approximation from cold to weakly-relativistic regimes of thermal plasma. In relativistic regime Faraday conversion largely deviated from the linear approximation, because it breaks at finite ratio of $\Omega_0/\omega$. Here $\Omega_0 = e B / (m_e c)$ is the cyclotron frequency and $\omega$ is the radiation frequency. Due to the lack of full investigation of various electron distributions, the important question was left unanswered: is thermal distribution special or such behavior of Faraday conversion is generic?

In present paper we expand the computations of Faraday conversion and rotation coefficients to non-thermal particle distributions.  We compute the absorption coefficients as well with the same unique method. We find solutions of the wave equation and natural modes from cold limit to ultra-relativistic limit. Our formulae are precise at all reasonable particle $\gamma$'s. We find a large discrepancy, if the linear approximations to Faraday conversion and rotation are used, thus justifying the need for precise computations.
To be practical we adopt $\delta$-function energy distribution of electrons and provide the fitting formulae, which can then be integrated over any isotropic distribution of particles. We also provide public code in Mathematica 8 to numerically compute the integrals. The paper is organized as follows. We derive dielectric tensor and dispersion relations for uniformly magnetized relativistic plasma with isotropic monoenergetic particle distribution in \S 2. The properties of natural modes are investigated in \S 3.  In \S 4 we provide simplified formulae and generalize to arbitrary electron energy distributions. on its polarization prediction is described In \S 5 we show that the polarized spectrum changes a lot, when linear approximations are used for Sgr A* modeling.

\section{RESPONSE TENSOR AND EIGENMODES OF UNIFORMLY MAGNETIZED RELATIVISTIC PLASMAS WITH $\delta$-FUNCTION ENERGY DISTRIBUTION}\label{disp_delta}
\subsection{Geometry and definitions}\label{geo}
\begin{figure}
\begin{center}
\includegraphics[width=70mm]{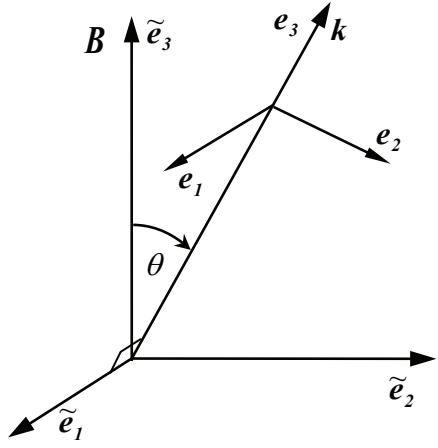}
\end{center}
\caption{Geometry of the problem. Vector ${\bmath B}$ represents uniform magnetic field. The transverse plane wave travels along $\bmath k$ and has electric field $\bmath E$ in $({\bmath e}_1, {\bmath e}_2)$ plane.}\label{figvector}
\end{figure}

Let us define a coordinate system in three-dimensional flat space with ${\bmath e}_1$ along the major axis of the synchrotron radiation ellipse,  ${\bmath e}_2$ along the minor axis, and ${\bmath e}_3$ towards the observer (see Figure \ref{figvector}). Vector ${\bmath e}_1$ is perpendicular to ${\bmath B}$. This coordinate system is right-handed, i.e., the observer finds a counter-clockwise rotation from ${\bmath e}_1$ to ${\bmath e}_2$. We define Minkowski space-time with basis $e^\mu_\alpha$, so that $e^\mu_0=(-1,0,0,0)$ and $e^\mu_j = (0, {\bmath e_j}), j=1,2,3$.  Unlike in Paper I, here we perform all derivations in spatial basis $( {\bmath e_1}, {\bmath e_2}, {\bmath e_3})$, instead of $(\tilde{\bmath e}_1, \tilde{\bmath e}_2, \tilde{\bmath e}_3)$. In latter basis $\tilde{\bmath e}_1={\bmath e}_1$ and $\tilde{\bmath e}_3 \parallel {\bmath B}$. We set the speed of light to unity $c=1$.

Normalized vectors of electric field in a transverse wave are
\begin{eqnarray} \label{E12}
	\hat{\bmath E}_1 &=& {\bmath e_1} \quad {\rm e}^{\imath (kX)}   \quad{\rm and} \quad \hat{\bmath E}_2 \quad=\quad {\bmath e_2} \quad {\rm e}^{\imath (kX)},
\end{eqnarray}
where $k_\mu=\omega(-1,0,0,1)^{\rm T}$ is the covariant photon momentum, $X^\mu$ are four-coordinates, and $kX$ represents the inner product $kX=k_\mu X^\mu$.
The projections of electric field ${\bmath E}$ of an arbitrary wave along these unit vectors have complex amplitudes $\widetilde{A}^{1,2}$, or real amplitudes $A^{1,2}$ and phases $\delta^{1,2}$, so that
\begin{eqnarray}\label{Eto12}
{\bmath E} = {\bmath E}_1 + {\bmath E}_2
	= \widetilde{A}^1 \hat{\bmath E}_1  + \widetilde{A}^2 \hat{\bmath E}_2 \quad=\quad A^1 {\rm e}^{i \delta^1} \hat{\bmath E}_1 + \quad A^2 {\rm e}^{i \delta^2} \hat{\bmath E}_2.
\end{eqnarray}
The tensor of intensity is defined as
\begin{eqnarray}
	{\mathcal I}_{i j} = <{\bmath E}_i \cdot conj[{\bmath E}_j]>,
\end{eqnarray}
where $conj[...]$ stands for complex conjugate and $<>$ represents the average over the wave ensemble.

Radiative transfer in uniform medium is described by the equation \citep{sazonov}
\begin{eqnarray}
\label{radtrans_tensor}
	\frac{\omega}{2\pi} \cdot \frac{\rm d}{{\rm d} s}  {\mathcal I}_{i j} &=& \varepsilon_{i j}  \quad+\quad  i  ( \alpha^{i m} {\mathcal I}_{m j} - \alpha^{*j n} {\mathcal I}_{i n} ),
\end{eqnarray}
where $s$ is the distance along the ray, $\varepsilon_{i j}$ is the tensor of spontaneous emission and $\alpha^{i j}$ is the tensor of wave propagation, or the response tensor.
The equation (\ref{radtrans_tensor}) can be rewritten in a more familiar form \citep{shcher_huang}
\begin{equation}\label{transfer}
\frac{d{\bmath S}}{ds}
=
\left(\begin{array}{c}
  \varepsilon_I \\  \varepsilon_Q \\  0 \\  \varepsilon_V
\end{array}\right)-
\left(%
\begin{array}{cccc}
  \eta_I & \eta_Q & 0 & \eta_V \\
  \eta_Q & \eta_I & \rho_V & 0 \\
  0 & -\rho_V & \eta_I & \rho_Q \\
  \eta_V & 0 & -\rho_Q & \eta_I \\
\end{array}%
\right)\bmath S
\end{equation} with the polarization vector
\begin{equation}\label{pol_vector}
{\bmath S}=(I,Q,U,V)^T
\end{equation} being the vector of Stokes parameters. The intensities ${\bmath S}$ and integrated polarized fluxes can be directly observed. Here $\varepsilon_I, \varepsilon_Q, \varepsilon_V$ are the emission coefficients,
\begin{eqnarray}\label{abs_def}
\eta_I&=&{\rm Im}(\alpha^{22}+\alpha^{11})/\nu,\nonumber\\
\eta_Q&=&{\rm Im}(\alpha^{11}-\alpha^{22})/\nu,\nonumber\\
\eta_V&=&2{\rm Re}(\alpha^{12})/\nu,
\end{eqnarray} are the absorption coefficients,
\begin{eqnarray}\label{prop_def}
\rho_V&=&2{\rm Im}(\alpha^{12})/\nu,\nonumber\\
\rho_Q&=&{\rm Re}(\alpha^{22}-\alpha^{11})/\nu,
\end{eqnarray} where $\nu=\omega/(2\pi)$ is the frequency. In the following we will concentrate mainly on Faraday rotation coefficient $\rho_V$ and Faraday conversion coefficient $\rho_Q$, which are generally called propagation coefficients. They directly influence the observed polarized fluxes.

\subsection{Response tensor and dispersion relations}\label{permit}
We start with the formula for the response tensor for isotropic electrons
\begin{eqnarray}
\label{resp0}
	\alpha^{\mu\nu}(k) =    - \frac{e^2}{m_e c}   \int {\rm d}^3 {\rm p} \frac{{\rm d} f(\gamma)}{{\rm d} \gamma}  \widetilde{\mathcal U}^\mu \widetilde{\mathcal U}^\nu      +    \frac{\imath e^2 \omega}{m_e c}       \int^\infty_0 {\rm d} \xi \dot{t}^\nu_\sigma (-\xi) \left[\frac{\partial^2}{\partial {\mathcal S}_\mu \partial {\mathcal S}_\sigma}  \int {\rm d}^3 {\rm p}    \left(  - \frac{1}{\gamma}  \frac{{\rm d} f(\gamma)}{{\rm d} \gamma} {\rm e}^{- \imath {\mathcal R}(\xi) {\mathcal U} + {\mathcal S} {\mathcal U}}  \right) \right]_{{\mathcal S}_\mu=0},
\end{eqnarray}
where $\widetilde{\mathcal U}^\mu=(1,0,0,0)$, ${\mathcal U}^\mu$ is 4-velocity of electrons in observer's Minkowskian frame, ${\rm p}$ is dimensionless 3-momentum defined as ${\rm p} = \sqrt{\gamma^2-1}$, and $f(\gamma)$ is the energy distribution function of electrons. It is normalized to the number density of electrons $n_e$ as
\begin{equation}
	\int f(\gamma) {\rm d^3 p}  = n_e.
\end{equation}
Tensor $\dot{t}^{\mu\nu}(\xi)$ describes how the velocity of electrons changes with proper time $\xi$, ${\mathcal R}(\xi) {\mathcal U}$ represents the difference of the phase ($kX$) of the electron, and ${\mathcal S}_\mu$ is an auxiliary variable.  The formula (\ref{resp0}) coincides with Eq.~(2.3.11), Chapter II, \citet{melrose2010} and with Eq. (19) in \citet{melrose97a}, except for $\dot{t}^{\mu\nu}(\xi)$, because we defined a different coordinate system.  Eq.~(35) in Paper I offers a similar expression derived in the 3-dimension space with basis $(\rm \tilde{\bmath e}_1, \tilde{\bmath e}_2, \tilde{\bmath e}_3)$.  The derivation of Eq.~\ref{resp0} and the related definitions can be
found in Appendices A \& B.

The 4-vectors in expression (\ref{resp0}) can be split into temporal and spatial parts as
\begin{equation}
{\mathcal R}_\mu  = (-\omega\xi, {\rm R}_j ), \quad {\mathcal U}^\mu = ( \gamma, {\rm u}^j ), \quad {\mathcal S}_\mu = ( s_0, {\rm s}_j ).
\end{equation} Then the phase becomes $-\imath {\mathcal R}(\xi) {\mathcal U} + {\mathcal S} {\mathcal U} =  \imath \omega\xi \cdot \gamma - \imath {\rm R}_j {\rm u}^j + s_0 \gamma + {\rm s}_j {\rm u}^j  =  \imath ( \omega \xi - \imath s_0 ) \gamma - \imath ( {\rm R}_j + \imath {\rm s}_j ) {\rm u}^j$.   The momentum integral in the response tensor is
\begin{eqnarray}
\label{respform}
	&\quad& \int {\rm d}^3 {\rm p}    \left(  - \frac{1}{\gamma}  \frac{{\rm d} f(\gamma)}{{\rm d} \gamma} {\rm e}^{ \imath ( \omega\xi  - \imath s_0 ) \gamma - \imath ( {\rm R}_j  + \imath {\rm s}_j  ) {\rm u}^j}  \right)   \nonumber \\
	&=& \int_{{\rm |p|}_{\rm min}}^{{\rm |p|}_{\rm max}}    {\rm d} {\rm |p|}  \cdot  {\rm |p|}^2  \cdot  \left( - \frac{1}{\gamma}  \frac{{\rm d} f(\gamma)}{{\rm d} \gamma} \right)  \cdot  {\rm e}^{\imath ( \omega \xi - i s_0 )  \gamma } \int_0^{2 \pi} \int_0^\pi  e^{- \imath  | {\rm\bf R} + \imath {\rm\bf s}| |{\rm\bf u}| \cos \theta_{\rm p} }  \sin \theta_{\rm p} {\rm d} \theta_{\rm p} {\rm d} \phi_{\rm p}  \nonumber \\
	&=& I (\xi, {\mathcal S}) -  \left[  f(\gamma) \cdot A(\gamma; \xi, \mathcal S)  \right]_{\gamma_{\rm min}}^{\gamma_{\rm max}} ,
\end{eqnarray}
where
\begin{eqnarray}
	I (\xi, {\mathcal S}) = 4\pi \cdot \int_{\gamma_{\rm min}}^{\gamma_{\rm max}}    f(\gamma) \cdot  e^{\imath ( \omega \xi - \imath s_0 ) \gamma}
\left[  \imath ( \omega \xi - \imath s_0 )  \cdot \frac{\sin \left( |{\rm\bf R} + \imath {\rm\bf s}| \sqrt{\gamma^2 -1} \right)}{|{\rm\bf R} + \imath {\rm\bf s}|}   \right.
	+ \left.  \frac{\gamma}{\sqrt{ \gamma^2 -1 }}  \cdot \cos \left( |{\rm\bf R} + \imath {\rm\bf s}| \sqrt{\gamma^2 -1} \right)  \right] {\rm d}\gamma     \nonumber
\end{eqnarray}
and
\begin{eqnarray}
	A (\gamma; \xi, {\mathcal S}) &=& 4\pi \cdot \frac{\sin ( |{\rm\bf R} + \imath {\rm\bf s}|  \sqrt{\gamma^2 -1} )}{|{\rm\bf R} + \imath {\rm\bf s}|}  \cdot  e^{\imath ( \omega \xi - \imath s_0 )  \gamma}.   \nonumber
\end{eqnarray}
The absolute values are taken as $|{\rm\bf R} + \imath {\rm\bf s}| = \sqrt{({\rm R}^k + \imath {\rm s}^k) ({\rm R}_k + \imath {\rm s}_k)}$ and $|{\rm\bf u}| = \sqrt{{\rm u}^k {\rm u}_k}$. Note that for any distribution with $ f(\gamma_{\rm min}) \cdot A(\gamma_{\rm min}) \to 0$ and $ f(\gamma_{\rm max}) \cdot A(\gamma_{\rm max}) \to 0$, the second term in last expression of Eq.~(\ref{respform})  vanishes.

As the first step of calculation for arbitrary distribution of electrons, we use a $\delta$-function as the distribution
\begin{equation}
f(\gamma)=\delta(\gamma - \gamma_0).
\end{equation}
Then $I(\xi, {\mathcal S})$ becomes $I(\gamma_0; \xi, {\mathcal S})$ as
\begin{eqnarray}
I(\gamma_0; \xi, {\mathcal S}) = {\rm e}^{\imath ( \omega \xi - \imath s_0 ) \gamma_0}  \left[  \imath ( \omega \xi  - \imath s_0 ) \cdot \frac{\sin \left( |{\rm\bf R} + \imath {\rm\bf s}| p_0 \right)}{|{\rm\bf R} + \imath {\rm\bf s}|}+  \frac{\gamma_0}{p_0}  \cdot \cos \left( |{\rm\bf R} + \imath {\rm\bf s}| p_0 \right)  \right],
\end{eqnarray} where
\begin{equation}
p_0=\sqrt{\gamma_0^2-1}
\end{equation} is the dimensionless momentum.

Now we apply the differential operator $\partial^2/(\partial {\mathcal S}_\mu \partial {\mathcal S}_\sigma)$ to $I(\gamma_0; \xi, {\mathcal S})$, set ${\mathcal S}_\mu=0$, and only choose $(\mu,\nu)=(1,2)$ to isolate the transverse wave component. The final expression of $2\times 2$ response tensor is
\begin{eqnarray}
\label{responsetensor}
\alpha^{ij}(k, \gamma_0) \quad=\quad   \frac{ 4\pi e^2}{m_e c}   \cdot \imath  \int^\infty_0 {\rm d} (\omega\xi) \cdot  {\rm e}^{\imath  \omega\xi \gamma_0}
\left[  \dot{t}^{ij} (\xi) \cdot ( \imath \omega\xi \Pi_1 + \Pi_3)  - \widetilde{T}^{ij}(\xi) \cdot ( \imath \omega\xi \Pi_2 + \Pi_4 ) \right],
\end{eqnarray}
where
\begin{eqnarray}
	\Pi_1 &=&    \frac{\sin \left( {\rm R(\xi)} p_0 \right)}{\rm R^3(\xi)} - p_0 \frac{\cos \left( {\rm R(\xi)} p_0 \right)}{\rm R^2(\xi)},   \nonumber \\
	 \Pi_2 &=&   \frac{ 3 \sin \left( {\rm R(\xi)} p_0 \right)}{\rm R^5(\xi)} - p_0 \frac{ 3 \cos \left( {\rm R(\xi)} p_0 \right)}{\rm R^4(\xi)}
	 - p_0^2 \frac{\sin \left( {\rm R(\xi)} p_0 \right)}{\rm R^3(\xi)},    \nonumber \\
	 \Pi_3 &=& \gamma_0 \frac{ \sin \left({\rm R}(\xi) p_0 \right) }{\rm R(\xi)},   \nonumber \\
	 \Pi_4 &=& \gamma_0    \frac{ \sin \left({\rm R}(\xi) p_0 \right) }{\rm R^3(\xi)}
	 - \gamma_0 p_0 \frac{ \cos \left({\rm R}(\xi) p_0 \right) }{\rm R^2(\xi)}\nonumber
\end{eqnarray}
and
\begin{eqnarray}
	{\rm R}(\xi) &=&  \sqrt{ {\rm R}^k(\xi) {\rm R}_k(\xi) }  \quad=\quad \sqrt{  \frac{\omega^2 \sin^2\theta}{\Omega_0^2} [ 2 - 2\cos(\Omega_0 \xi) ] + \cos^2\theta \cdot \omega^2 \xi^2 },  \nonumber \\
	\dot{t}^{ij}(\xi) &=& \left( \begin{array}{cc}
		\cos(\Omega_0 \xi) & - \cos\theta \sin(\Omega_0 \xi) \\
		\cos\theta \cdot\sin(\Omega_0 \xi) & \sin^2\theta + \cos^2\theta \cdot\cos(\Omega_0 \xi) \end{array} \right),  \nonumber \\
	\widetilde{T}^{ij}(\xi) &=& {\rm R}^i (\xi) \widetilde{{\rm R}}^j (\xi) \quad=\quad \frac{\omega^2\sin^2\theta}{\Omega_0^2}    \left( \begin{array}{cc}
		-  \left( 1 - \cos(\Omega_0 \xi) \right)^2 & - \cos\theta \left( \sin(\Omega_0 \xi) - \Omega_0 \xi \right) \left( 1 - \cos(\Omega_0 \xi) \right)  \\
		\cos\theta \left( \sin(\Omega_0 \xi) - \Omega_0 \xi \right) \left( 1 - \cos(\Omega_0 \xi) \right) & \cos^2\theta \left( \sin(\Omega_0 \xi) - \Omega_0 \xi \right)^2 \end{array} \right). \nonumber
\end{eqnarray} The differentiation of $A(\gamma; \xi, {\mathcal S})$ yields a response tensor boundary term as
\begin{eqnarray}
\label{responsetensorB}
	\alpha_{\rm B}^{ij}(k, \gamma_{\rm min; max}) &=&   \frac{ 4\pi  e^2 }{m_e c}   \cdot \imath \int^\infty_0 {\rm d} (\omega \xi)   {\rm e}^{\imath  \omega \xi \gamma_{\rm min; max}}  \left[  \dot{t}^{ij} (\xi) \cdot \Pi_1  - \widetilde{T}^{ij}(\xi)  \cdot \Pi_2 \right].
\end{eqnarray} The expression (\ref{responsetensorB}) is needed for distributions confined by cut-off Lorentz factors.

The dielectric tensor
\begin{eqnarray}
	\varepsilon^{i j} &=& \delta^{i j} + \frac{4\pi c}{\omega^2} \alpha^{i j}
\end{eqnarray} leads to the wave equation
\begin{eqnarray}
	\left(\frac{k^2 c^2}{\omega^2} \delta^{i j} - \varepsilon^{i j}\right)
	 \left( \begin{array}{cc}
		\hat{E}_1 \\ \hat{E}_2 \\
	\end{array} \right) &=& 0.
\end{eqnarray}
The corresponding wave dispersion relation is
\begin{eqnarray}
	k_\pm^2 c^2 &=&  \omega^2 + 2 \pi c \left[ \alpha^{11} + \alpha^{22} \pm \sqrt{ (\alpha^{11} - \alpha^{22})^2 + 4\alpha^{12}\alpha^{21} } \right],
\end{eqnarray}
where $\alpha^{21} = - \alpha^{12}$ due to Onsager principle \citep{landau10} (p. 273).
Note that $k_\pm^2$ are real, when $\alpha^{i j}$ only has the Hermitian part, while $k_\pm^2$ become complex, if $\alpha^{i j}$ has both the Hermitian and the anti-Hermitian parts. The approximate relation $k_\pm^2 c^2\approx\omega^2 + (2\pi c){\rm Re}(\alpha^{11} + \alpha^{22})$ helps to determine if the waves can propagate through medium.
For example, the number density $n_e < 10^7$ was estimated near Sgr A* \citep{yuanf03}, so that $k_\pm$ are dominated by their real parts for $\omega > 100$MHz and the waves can propagate.

\section{NUMERICAL CALCULATION OF RESPONSE TENSOR AND THE ELLIPTICAL NATURE OF NATURAL MODES}\label{numcal}
\subsection{Numerical calculation of response tensor}

We substitute the components of the response tensor from Eq.~(\ref{responsetensor}) and perform integration over $\xi$ in complex plane in Mathematica 8.
The source code in Mathematica 8 can be found at \url{http://astroman.org/Faraday_conversion/}. We then find propagation coefficients $\rho_V$ and $\rho_Q$ and absorption coefficients $\eta_I,$ $\eta_Q,$ and $\eta_V$ according to relations (\ref{abs_def},\ref{prop_def}). We analogously compute the boundary terms by substituting the components of $\alpha^{ij}_{\rm B}$ from Eq.~(\ref{responsetensorB}). Just like in Paper I, we do not perform the integration over $\xi$ along the real axis. To accelerate the convergence we integrate in a complex plane along the ray originating at $\xi=0$ at a positive angle $\psi \in (0, \pi/2)$ to the real axis. Angle $\psi$ needs to be small enough in order to avoid crossing the branch points of $\alpha^{ij}$. These branch points are produced by zeros of $R(\xi)$.  We integrate the full complex expressions to find the response tensor $\alpha^{ij}$ and the boundary term. If $\gamma_0$ is small, then computations of anti-Hermitian parts involve substantial cancelations with the values of integrals close to zero. Thus it is hard to reach good accuracy for absorptivity calculations with the method chosen, whereas the correspondent Hermitian parts easily converge. When $\gamma_0$ is larger than $\sim 10$, the values of Hermitian-related part and anti-Hermitian-related part become comparable and all integrals converge.

\begin{figure}
\begin{center}
\includegraphics[scale=1.1]{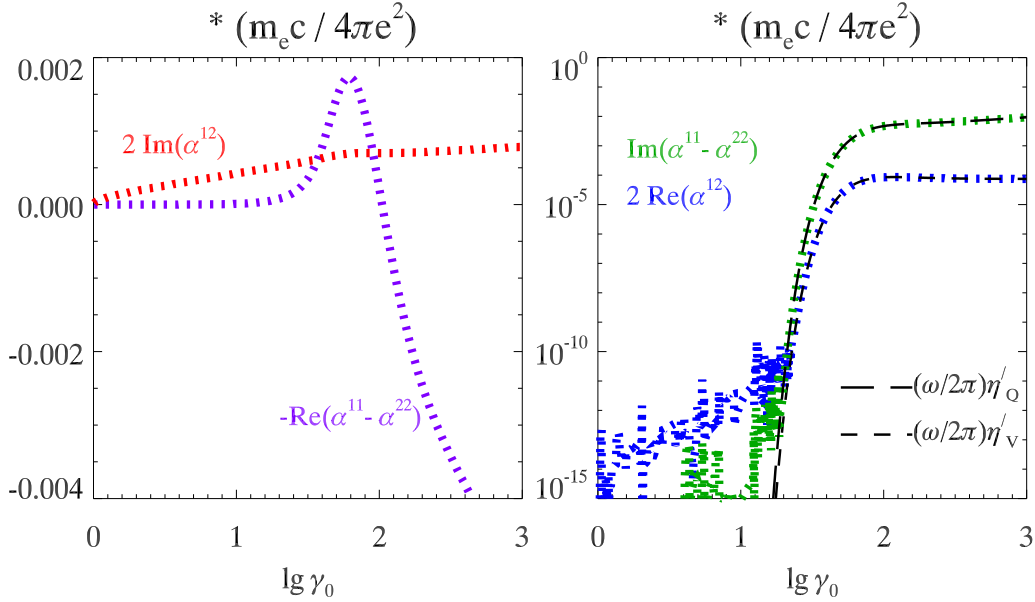}
\end{center}
\caption{Absorption and propagation coefficients for the fiducial model with parameters $\Omega_0/\omega = 10^{-4}$ and $\theta=45^\circ$ for the monoenergetic electron distribution. {\it Left}:  $2 {\rm Im}(\alpha^{12})\propto \rho_V$ (Faraday rotation, in red) and $- {\rm Re}(\alpha^{11}-\alpha^{22})\propto \rho_Q$ (Faraday conversion, in purple).   {\it Right}:  ${\rm Im}(\alpha^{11}-\alpha^{22})\propto\eta_Q$ (absorption of linearly-polarized waves, green) and $2 {\rm Re}(\alpha^{12})\propto\eta_V$ (absorption of circularly-polarized waves, blue).   Approximate absorption coefficients based on \citet{sazonov}, in particular, $(\omega/2\pi) \eta'_Q$ and $(\omega/2\pi) \eta'_V$, are shown in long-dashed and dashed black lines, respectively.}\label{figdelta}
\end{figure}

We choose the fiducial model with $\Omega_0 /\omega = 10^{-4}$ and $\theta = 45^\circ$ and plot on Figure~\ref{figdelta} propagation coefficients and absorption coefficients as functions of the Lorentz factor. On the left panel, $2 {\rm Im}(\alpha^{12})\propto \rho_V$ and $- {\rm Re}(\alpha^{11}-\alpha^{22})\propto \rho_Q$, multiplied by $(m_e c) / (4\pi e^2)$, are shown in red and purple dotted lines, respectively.  They both monotonically increase as $\gamma_0$ increases from 1 to $\sim 60$. As $\gamma_0$ increases further, the profile of $\rho_V$ becomes flatter, while $\rho_Q$ reaches its peak and decreases to negative values.

On the right panel, ${\rm Im}(\alpha^{11}-\alpha^{22})\propto\eta_Q$ and $2 {\rm Re}(\alpha^{12})\propto\eta_V$, divided by $(4\pi e^2 /m_e c)$, are shown in green and blue dotted lines, respectively.  We also show $(\omega/2\pi) \eta'_Q$ and $(\omega/2\pi) \eta'_V$ as long-dashed and dashed black lines, respectively.  Here $\eta'_Q$ and $\eta'_V$ corresponds to the integrals in \citet{sazonov}\footnote{The sign of $\eta_V$ is opposite to that in \citet{sazonov}, if the current IAU/IEEE definition of the sign of circular polarization is followed.}
\begin{eqnarray}
\label{sazonovab}
	\eta'_Q &=& - \frac{\nu_B \sin\theta}{\nu} \frac{\sqrt{3} e^2}{4 m_e c \nu}
\int_1^\infty {\rm d} \gamma   \cdot \gamma^2   \frac{\partial}{\partial \gamma} \left[  \frac{\delta(\gamma - \gamma_0)}{\gamma^2}  \right]  \frac{\nu}{\nu_\gamma} K_{2/3} \left(\frac{\nu}{\nu_\gamma} \right)   \nonumber \\
	\eta'_V &=& - \frac{\nu_B \cos\theta}{\nu} \frac{e^2}{\sqrt{3} m_e c \nu}   \int_1^\infty {\rm d} \gamma   \cdot \gamma   \frac{\partial}{\partial \gamma} \left[  \frac{\delta(\gamma - \gamma_0)}{\gamma^2}  \right]
	\left[ \frac{\nu}{\nu_\gamma} K_{1/3} \left(\frac{\nu}{\nu_\gamma} \right)  +  \int_{\nu/\nu_\gamma}^\infty {\rm d z} K_{1/3} (z)  \right],
\end{eqnarray}
where $\nu_B$ is the cyclotron frequency, $\nu_\gamma = (3 e B \sin\theta \gamma^2)/(4\pi m_e c) $ is the characteristic frequency, and $K_\alpha$ is the modified Bessel function of the second kind of order $\alpha$.   We integrate by parts to deal with the differential of the $\delta$-function. Note that our absorptivities deviate a lot from the correct values, when $\gamma_0$ is small. As mentioned earlier, this is due to large cancelations of the parts of the integral, so that absorptivities cannot easily converge.  When $\gamma_0 > 30$, they coincide with the approximate expressions.

Despite inaccuracy at low Lorentz factors, the calculation clearly shows an important property of plasma absorption.  The absorption in $Q$-component is smaller, compared to that in $V$-component $\eta_Q<|\eta_V|$ at low Lorentz factors, while $\eta_Q>|\eta_V|$ at high Lorentz factors.  This shows that the radiation mechanism changes from CP-dominated cyclotron to LP-dominated synchrotron as the Lorentz factor increases.  The traditional approximations do not exhibit this property, because they assume high Lorentz factors.

\subsection{The axial radios and natural modes}\label{natmodes}
The corresponding eigenvectors from the wave equation are $(\widetilde{T}^+, \imath)^{\rm T}$ and $(\widetilde{T}^-, \imath)^{\rm T}$, respectively, where
\begin{eqnarray}
	\widetilde{T}^+ &=& \frac{\alpha^{11} - \alpha^{22} + \sqrt{ (\alpha^{11} - \alpha^{22})^2 + 4\alpha^{12}\alpha^{21} } }{2 \imath \alpha^{12}},  \nonumber \\
	\widetilde{T}^- &=& \frac{\alpha^{11} - \alpha^{22} - \sqrt{ (\alpha^{11} - \alpha^{22})^2 + 4\alpha^{12}\alpha^{21} } }{2 \imath \alpha^{12}}.
\end{eqnarray}

These $\widetilde{T}^+$ and $\widetilde{T}^-$ are complex axial ratios of the transverse wave. They obey the relation
\begin{eqnarray}
	\widetilde{T}^+ \widetilde{T}^- &=& -1.
\end{eqnarray}
The polar decomposition into real amplitudes $T^\pm (\ge 0)$ and phases $\varphi^\pm$ reads
\begin{eqnarray}
	\widetilde{T}^+ &=& T^+ {\rm e}^{\imath \varphi^+}  \quad{\rm and}\quad \widetilde{T}^- \quad=\quad T^- {\rm e}^{\imath \varphi^-},
\end{eqnarray}
so that
\begin{eqnarray}
	T^+ T^- &=& 1  \quad{\rm and}\quad \varphi^+ + \varphi^-  \quad=\quad - \pi.
\end{eqnarray}
These two wave eigenvalues define two natural wave modes as
\begin{eqnarray}
	\hat{\bmath E}_+ &=& \frac{\widetilde{T}^+ \hat{\bmath E}_1 + \imath \hat{\bmath E}_2}{\sqrt{1 + \widetilde{T}^+ \cdot conj[\widetilde{T}^+]}} \quad=\quad \frac{T^+ {\rm e}^{\imath \varphi^+} \hat{\bmath E}_1 + \imath \hat{\bmath E}_2}{\sqrt{1 + \left(T^+\right)^2}}  \nonumber \\
	\hat{\bmath E}_- &=& \frac{\widetilde{T}^- \hat{\bmath E}_1 + \imath \hat{\bmath E}_2}{\sqrt{1 + \widetilde{T}^- \cdot  conj[\widetilde{T}^-}]} \quad=\quad \frac{T^- {\rm e}^{\imath \varphi^-} \hat{\bmath E}_1 + \imath \hat{\bmath E}_2}{\sqrt{1 + \left(T^-\right)^2}}  
\end{eqnarray}
In practise, $\varphi^+ <  - \pi/2$ and $\varphi^- > - \pi/2$. Thus, the electric vector $\hat{\bmath E}_+$ rotates counter-clockwise and the electric vector $\hat{\bmath E}_-$ clockwise, as seen by the observer.
Note that
\begin{eqnarray}
	\hat{\bmath E}_+ \cdot conj[\hat{\bmath E}_-] &=& \frac{{\rm e}^{\imath (\varphi^+ - \varphi^-)} + 1}{T^+ + T^-}  \quad \left\{
	\begin{array}{cc}
		= 0,  & {\rm iff}\quad \varphi^+ = -\pi \\
		\ne 0, & {\rm otherwise}
	\end{array} \right.
\end{eqnarray}
That is $\hat{\bmath E}_+$ and $\hat{\bmath E}_-$ are not perpendicular to each other, unless the anti-Hermitian part, absorption, can be neglected.

\subsection{Properties of radiation in natural modes}\label{propnatmode}
\begin{figure}\begin{center}
\includegraphics[scale = 1.1]{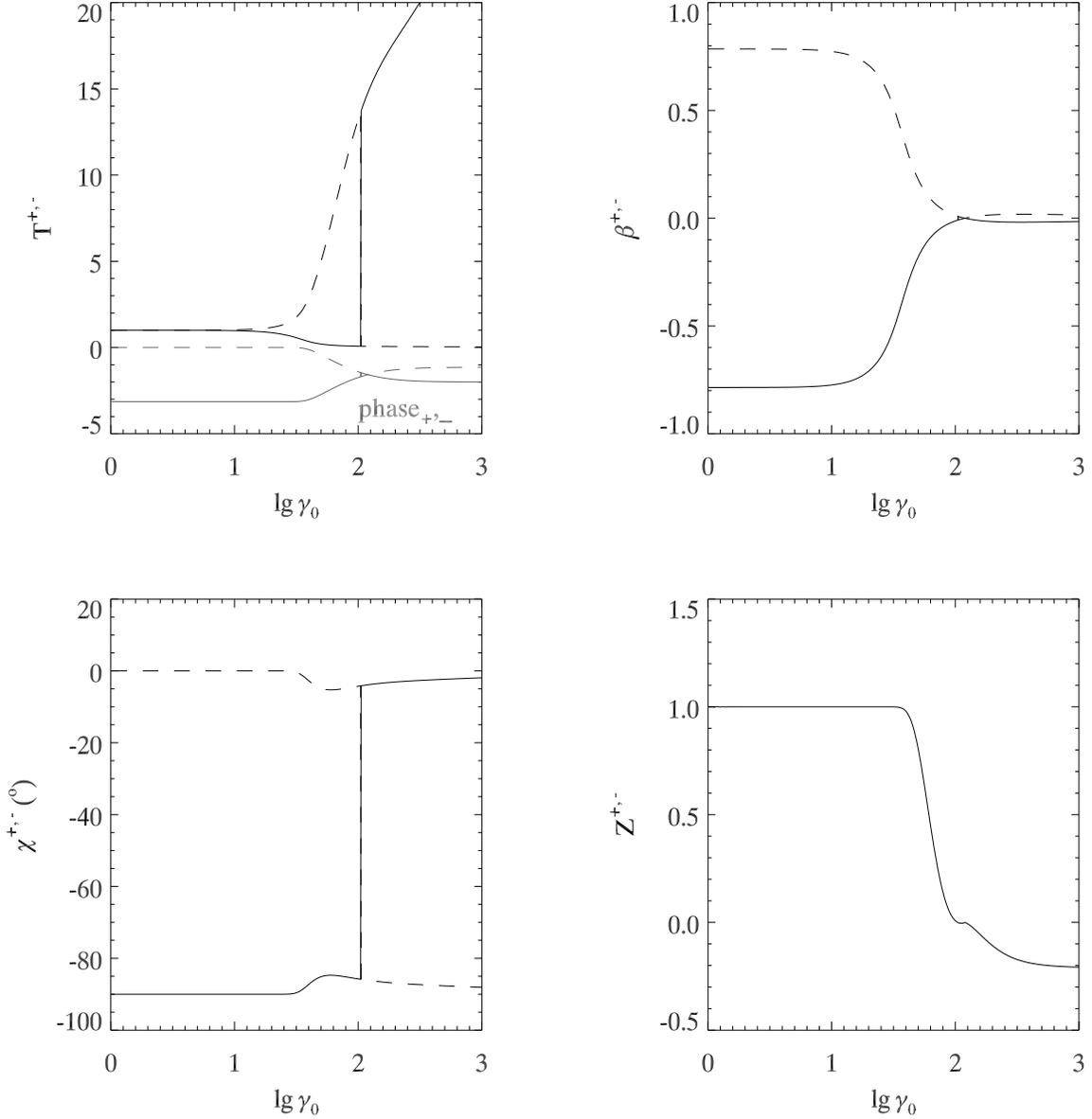}
\end{center}
\caption{Properties of two natural modes for the monoenergetic particle distribution. Amplitudes $T^\pm$ (black) and phases $\varphi^\pm$  (grey) of complex axial ratio, ellipticity $\beta^\pm$ ($\sim \pm \pi/4$ corresponding to circular polarized waves and $\sim 0$ to linear ones),  EVPA $\chi^\pm$, and $Z^+=Z^-$ (eq.~\ref{Zmonster}) are shown in four panels, respectively. Lines for $(^+)$-mode are solid and for $(^-)$-mode are dashed.}\label{figmodedelta}
\end{figure}

Let us decompose the natural modes along $\hat{\bmath E}_1$ and $\hat{\bmath E}_2$ as
\begin{eqnarray}
	{\bmath E}^\pm_1 &=& \frac{A^\pm  T^\pm {\rm e}^{\imath \varphi^+}}{\sqrt{1 + \left(T^\pm\right)^2}} \hat{\bmath E}_1  \quad{\rm and}\quad {\bmath E}^\pm_2 \quad=\quad \frac{\imath  A^\pm }{\sqrt{1 + \left(T^\pm\right)^2}} \hat{\bmath E}_2
\end{eqnarray}
We define the Stokes parameters for natural modes as
\begin{eqnarray}
	I_0^\pm &=& <{\bmath E}^\pm_1 \cdot conj[{\bmath E}^\pm_1]> + <{\bmath E}^\pm_2 \cdot conj[{\bmath E}^\pm_2]> = \left(A^\pm\right)^2,   \nonumber \\
	Q_0^\pm &=& <{\bmath E}^\pm_1 \cdot conj[{\bmath E}^\pm_1]> - <{\bmath E}^\pm_2 \cdot conj[{\bmath E}^\pm_2]> = \left(A^\pm\right)^2 \cdot \frac{\left(T^\pm\right)^2 - 1}{1 + \left(T^\pm\right)^2}, \nonumber \\
	U_0^\pm &=& <{\bmath E}^\pm_1 \cdot conj[{\bmath E}^\pm_2]> + <{\bmath E}^\pm_2 \cdot conj[{\bmath E}^\pm_1]> = \left(A^\pm\right)^2 \cdot  \frac{2 T^\pm \sin \varphi^\pm}{1 + \left(T^\pm\right)^2}, \nonumber \\
	V_0^\pm &=& \imath \left( <{\bmath E}^\pm_1 \cdot conj[{\bmath E}^\pm_2]> - <{\bmath E}^\pm_2 \cdot conj[{\bmath E}^\pm_1]> \right)= \left(A^\pm\right)^2 \cdot  \frac{2 T^\pm \cos \varphi^\pm}{1 + \left(T^\pm\right)^2}.
\end{eqnarray}
Note that $X_0 \ne X_0^+ + X_0^-,  (X_0=I_0,Q_0,U_0,V_0)$, since $<\hat{\bmath E}_+ \cdot conj[\hat{\bmath E}_-]> \ne 0$ unless $\varphi^+ = -\pi$.
These Stokes parameters correspond to elliptically polarized radiation with ellipticity $\beta$
\begin{eqnarray}
	\beta^\pm &=& \frac{1}{2} \sin^{-1} \frac{V_0^\pm}{I_0^\pm} \quad=\quad \frac{1}{2} \sin^{-1} \left( \frac{2 T^\pm \cos \varphi^\pm}{1 + \left(T^\pm\right)^2} \right)
\end{eqnarray}
and the electric vector position angle (EVPA)
\begin{eqnarray}
	\chi^\pm &=& \frac{1}{2} \tan^{-1} \frac{U_0^\pm}{Q_0^\pm} \quad=\quad \frac{1}{2} \tan^{-1} \left( \frac{2 T^\pm \sin \varphi^\pm}{\left(T^\pm\right)^2 - 1} \right).
\end{eqnarray}
The relations between the axial ratios and the Stokes parameters for each mode are
\begin{eqnarray}
\frac{\left(R^+_{\rm CP}\right)^2}{\left(R^+_{\rm LP}\right)^2} \left( = \frac{\left(V_0^+\right)^2}{\left(Q_0^+\right)^2 + \left(U_0^+\right)^2} \right)  \quad=\quad \frac{\left(R^-_{\rm CP}\right)^2}{\left(R^-_{\rm LP}\right)^2} \left( = \frac{\left(V_0^-\right)^2}{\left(Q_0^-\right)^2 + \left(U_0^-\right)^2} \right)= \left(  \frac{2}{|\widetilde{T}^+ + \widetilde{T}^-|} \right)^2  |  {\rm Re}( \widetilde{T}^+ ) {\rm Re}( \widetilde{T}^- ) |.
\end{eqnarray}
We define a special quantity
\begin{equation}\label{Zmonster}
Z^\pm=\frac{R^{\pm}_{\rm CP}}{R^{\pm}_{\rm LP}} \cdot  \frac{\rho_Q}{\rho_V},
\end{equation} which effectively measures how well the normal modes can be described, if we ignore absorption. We show amplitudes $T^\pm (\ge 0)$, phases $\varphi^\pm$, ellipticities $\beta^\pm$, and EVPA $\chi^\pm$ on Figure \ref{figmodedelta}. All quantities with superscript $(^+)$ are drawn in solid lines and those with $(^-)$ are dashed.

\subsubsection{Cold plasma limit}
In cold plasma the anti-Hermitian part of the response tensor can be neglected compared to the Hermitian part.  In this case, $\varphi^+ = - \pi$, $\varphi^- = 0 $ (see grey lines on the top left panel of Figure~\ref{figmodedelta}),  $U_0^+ = U_0^- = 0$, and $|\rho_V| >> |\rho_Q|$. The axial ratios are
\begin{eqnarray}
	\widetilde{T}^+ &=&  \frac{1 - \sqrt{ 1  + ( \rho_V/\rho_Q )^2}}{\rho_V/\rho_Q} \quad=\quad - T^+ \quad\approx\quad -1,  \nonumber \\
	 \widetilde{T}^- &=&  \frac{1 + \sqrt{ 1  + ( \rho_V/\rho_Q )^2}}{\rho_V/\rho_Q} \quad=\quad T^- \quad\approx\quad 1.
\end{eqnarray}
They are consistent with the Eq.(4.6) in \citet{melrose97b}.  The corresponding ellipticity $\beta$ and EVPAs are
\begin{eqnarray}
	\beta^+ &=& - \beta^- \quad=\quad \frac{1}{2} \sin^{-1} \left( - \frac{2}{T^+ + T^-} \right)  \quad\approx\quad  - \frac{\pi}{4}, \nonumber \\
	\chi^+ &=& - \frac{\pi}{2}, \quad \chi^- \quad=\quad 0.
\end{eqnarray}
Two natural modes are circularly polarized. They are orthogonal with major and minor axes aligned with ${\bmath e}_1$ and ${\bmath e}_2$, respectively. The relation between the Stokes parameters and the axial ratios becomes
\begin{eqnarray}
	\frac{R^\pm_{\rm CP}}{R^\pm_{\rm LP}} = \frac{R_{\rm CP}}{R_{\rm LP}} \quad=\quad \frac{V_0}{Q_0} \quad=\quad \frac{2}{T^+ - T^-} \quad=\quad \frac{\rho_V}{\rho_Q} , \quad
	{\rm i.e.,} \quad Z^\pm \quad=\quad 1.
\end{eqnarray}
This means the ratio of circular to linear radiation intensities in cold plasma eigenmodes equals the ratio of Faraday rotation to Faraday conversion coefficients. The total emission is dispersionless in a sense that the term $(\vec{\rho} \times \vec{p})$ in transfer of polarized radiation in \citet{L_L} vanishes, also see \citet{huang09a} for similar results with different definitions of the axial ratios. As Faraday rotation is much stronger than Faraday conversion, then the circular polarized intensity is much larger than the linear polarized intensity in eigenmodes. These are the well-known properties of 'cyclotron'/cold plasma regime.

In the limit $\omega^2 >> \alpha^{\mu\nu}$, the refractive indices of two modes are
\begin{eqnarray}
	n_\pm = \frac{k_\pm c}{\omega} \quad=\quad \sqrt{ 1 + \frac{2 \pi c}{\omega^2} ( \alpha^{11} + \alpha^{22} \pm \frac{\omega}{2 \pi} \rho_V ) }
	\approx 1 + \frac{1}{2} \frac{2 \pi c}{\omega^2} ( \alpha^{11} + \alpha^{22} \pm \frac{\omega}{2 \pi} \rho_V ),
\end{eqnarray}
which corresponds to standard rotation by $\theta_f$ of EVPA plane, Faraday rotation effect, as
\begin{eqnarray}
	\frac{ {\rm d}\theta_f }{{\rm d} l} &=& \frac{\omega}{2 c} (n_+ - n_-)  \quad\approx\quad \frac{1}{2} \rho_V
\end{eqnarray}
\subsubsection{Ultrarelativistic plasma limit}
In plasma with very high Lorentz factors $\rho_Q$ changes its sign to the negative. Two natural modes also change their polarization property from circular to linear. We find an interesting result that the quantity $Z^\pm$ is approaching another constant.  This means the proportion of linear to circular radiation in each eigenmode can still be easily measured by the proportion of Faraday conversion to rotation.

As it appears $\eta_Q \approx - 2 \rho_Q$ in this limit, and $|\eta_V| << |\rho_V| << |\rho_Q|$. So the complex axial ratios become
\begin{eqnarray}
	&\quad& \widetilde{T}^\pm \quad=\quad  \frac{\rho_Q}{\rho_V} \frac{-1 + \imath \left(\frac{\eta_Q}{\rho_Q}\right) \mp \sqrt{\left[-1 + \imath \left(\frac{\eta_Q}{\rho_Q}\right)\right]^2 - \left[\left(\frac{\eta_V}{\rho_Q}\right) + \imath \left(\frac{\rho_V}{\rho_Q}\right)\right]^2}}{ \imath \left[\left(\frac{\eta_V}{\rho_V}\right) + \imath \right]}  \nonumber\\
	&\approx& \frac{\rho_Q}{\rho_V}  \left[  1 + 2 \imath  \pm \sqrt{  ( 1 + 2 \imath )^2  + (\rho_V /\rho_Q)^2 }  \right]
	\approx \frac{\rho_Q}{\rho_V}  \left[  1 + 2 \imath  \pm (1 + 2 \imath) \left(1 + \frac{1}{2} \left( \frac{\rho_V/\rho_Q}{1 + 2\imath} \right)^2 \right)   \right].
\end{eqnarray}
We derive
\begin{eqnarray}
	\widetilde{T}^+ &\to& \frac{\rho_Q}{\rho_V} \cdot (2 + 4 \imath), \quad T^+ \to \infty, \quad \varphi^+ \to -\pi + \tan^{-1} (2) \approx - 2, \nonumber \\
	\widetilde{T}^- &\to& - \frac{\rho_V}{\rho_Q} \cdot (0.1 - 0.2 \imath), \quad T^- \to 0, \quad \varphi^- = -\pi - \varphi^+ \to - \tan^{-1} (2) \approx  - 1, \nonumber \\
	\beta^\pm &\to& 0,  \quad \chi^+ \to  0, \quad \chi^- \quad\to\quad - \frac{\pi}{2},
\end{eqnarray}
and
\begin{eqnarray}
	&\quad& \frac{R^{+}_{\rm CP}}{R^{+}_{\rm LP}} \quad=\quad \frac{R^{-}_{\rm CP}}{R^{-}_{\rm LP}}
	=\frac{2}{|\widetilde{T}^+ + \widetilde{T}^-|}  \sqrt{| {\rm Re}(\widetilde{T}^+) {\rm Re}(\widetilde{T}^-) |}
	\to  \frac{2}{4.47 \cdot |\rho_Q / \rho_V|} \sqrt{ 2 \frac{\rho_Q}{\rho_V} \cdot 0.1 \frac{\rho_V}{\rho_Q} }\approx - 0.2 \cdot \frac{\rho_V}{\rho_Q}  \nonumber \\
	&{\rm i.e.,}& \quad Z^\pm \quad\to\quad - 0.2
\end{eqnarray}
Note that the total intensity cannot be calculated by simply adding intensities in two modes, because they are not orthogonal.

\subsubsection{Intermediate regime}
In plasma with intermediate Lorentz factors (about $\gamma_0 \sim 10^{1-2}$ for the fiducial model with $\Omega_0/\omega=10^{-4}$), the properties of the natural modes change gradually from those in cold limit to those in ultra-relativistic limit.

As $\gamma_0$ increases, $T^+$ decreases to $0$, while $T^-$ approaches $\infty$, ellipticities $|\beta^\pm|$ decrease to $0$, EVPAs $\chi^\pm$ deviate from $-90^\circ$ (or $0^\circ$).  At a special value of $\gamma_0^*$ the Faraday conversion coefficient $-{\rm Re} (\alpha^{11} - \alpha^{22})$ changes its sign, as shown in Figure~\ref{figdelta}. The Lorentz factor for the fiducial model is $\gamma_0^*\approx 100$. Flips of $T^\pm$, $\beta^\pm$, and $\chi^\pm$ between two modes help to preserve the handedness of the modes. That is the wave with $\hat{\bmath E}_+$ always has a counter-clockwise rotation, while the wave described by $\hat{\bmath E}_-$ has a clockwise one.  The value of $Z^\pm$ decreases in the intermediate regime from $1$ to $\sim -0.2$.

\section{PROPAGATION AND ABSORPTION COEFFICIENTS FOR ISOTROPIC PLASMAS WITH ARBITRARY ENERGY DISTRIBUTIONS}\label{arbitdis}
\subsection{Integration over particle distribution function}\label{energint}
The final form of the response tensor is related to $I(\xi, {\mathcal S})$ as shown by Eq.~(\ref{respform}).  With the aid of
\begin{eqnarray}
	I(\xi, {\mathcal S}) &=&\int f(\gamma) I(\gamma_0; \xi, {\mathcal S}) {\rm d}\gamma,
\end{eqnarray}
one has
\begin{eqnarray}
	\alpha^{ij}(k) &=&  \left\{ \frac{\imath e^2 \omega}{m_e c}       \int^\infty_0 {\rm d} \xi \dot{t}^\nu_\sigma (-\xi) \left[ \frac{\partial^2}{\partial {\mathcal S}_\mu \partial {\mathcal S}_\sigma}  I (\xi, {\mathcal S})    -  \left(  f(\gamma) \cdot \frac{\partial^2}{\partial {\mathcal S}_\mu \partial {\mathcal S}_\sigma} A(\gamma; \xi, \mathcal S)  \right)_{\gamma_{\rm min}}^{\gamma_{\rm max}}   \right]_{s_\mu=0}  \right\}_{\mu,\nu=1,2}   \nonumber \\
	&=& \int_{\gamma_{\rm min}}^{\gamma^{\rm max}} f(\gamma) \cdot \alpha^{ij}(k, \gamma) {\rm d} \gamma
-  f(\gamma_{\rm max}) \cdot \alpha^{ij}_{\rm B}(k, \gamma_{\rm max}) + f(\gamma_{\rm min}) \cdot \alpha^{ij}_{\rm B}(k, \gamma_{\rm min}).
\end{eqnarray}
The isotropic distribution function $f(\gamma)$ is normalized as $\int f(\gamma) {\rm d^3p}=n_e$.   In this section we compute the propagation and absorption coefficients for thermal and power-law particle distributions.

\subsubsection{Thermal distribution}\label{arbitthermal}
\begin{figure}
\begin{center}
\includegraphics[scale = 1.1]{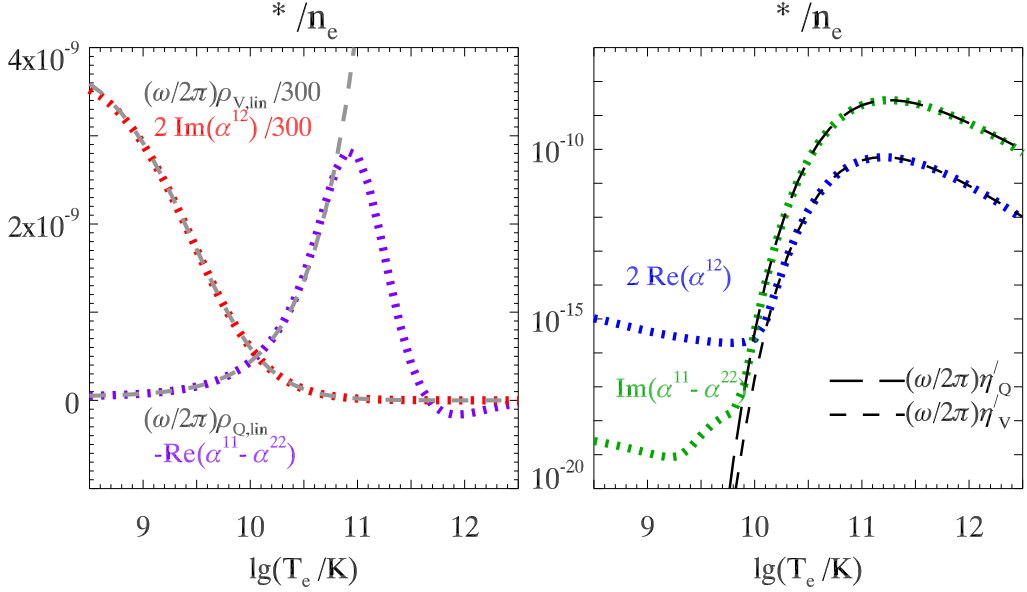}
\end{center}
\caption{Propagation and absorption coefficients for thermal energy distribution.  {\it Left}: Faraday rotation and conversion. Linear approximations by \citet{melrose97b} are shown in black dashed lines.   {\it Right}: Absorption coefficients. Traditional approximations are shown in black solid lines.}\label{figthermal}
\end{figure}

The thermal distribution is
\begin{eqnarray}
	f(\gamma) &=& \frac{n_e}{4\pi \Theta_e K_2(\Theta_e^{-1})} {\rm e}^{- \gamma/\Theta_e}, \quad 1\le\gamma<+\infty,
\end{eqnarray}
where $\Theta_e = {\rm k_B} T_e /(m_e c^2)$ is the dimensionless particle temperature. It is normalized to number density of electrons $n_e$  as
\begin{eqnarray}
	\int f(\gamma) {\rm d^3 p} = 4\pi  \int_1^{+\infty} \gamma \sqrt{\gamma^2-1} f(\gamma) {\rm d} \gamma = n_e.
\end{eqnarray}
In this case, $f(\gamma_{\rm max} \to +\infty) \to 0$, so that the response tensor becomes
\begin{eqnarray}
	\alpha^{ij}(k) = \frac{n_e}{4\pi \Theta_e K_2(\Theta_e^{-1})}  \left[  \int_1^{+\infty}  {\rm e}^{- \gamma/\Theta_e}\cdot \alpha^{ij}(k, \gamma) {\rm d} \gamma
+ {\rm e}^{- 1/\Theta_e}\cdot \alpha_{\rm B}^{ij}(k, 1)  \right]
	= \frac{n_e}{4\pi \Theta_e K_2(\Theta_e^{-1})}  \int_1^{+\infty}  {\rm e}^{- \gamma/\Theta_e}\cdot \alpha^{ij}(k, \gamma) {\rm d} \gamma,
\end{eqnarray}
since $\alpha_{\rm B}^{ij}(k, 1) \to 0$.  The propagation and absorption coefficients for $n_e=1$ are shown in thick dotted lines on Figure~\ref{figthermal}.

On the left panel we show in dashed grey lines the linear approximations to propagation coefficients elaborated in \citet{melrose97b,shcher_farad} for $n_e=1$. The related formulae are\footnote{Note, that Faraday conversion coefficient in the "linear" regime is actually proportional to $(\Omega_0/\omega)^2$.}
\begin{eqnarray}\label{lin_approx}
	\rho_{Q, {\rm lin}} &=& \frac{2\pi e^2 \Omega_0^2}{m_e c \cdot \omega^3} \left( \frac{K_1(\Theta_e^{-1})}{K_2(\Theta_e^{-1})} \sin^2 \theta + 6 \Theta_e \sin^2 \theta
\right),   \nonumber\\ \rho_{V, {\rm lin}} &=& \frac{4\pi e^2 \Omega_0}{m_e c \cdot \omega^2} \frac{K_0(\Theta_e^{-1})}{K_2(\Theta_e^{-1})} \cos \theta.
\end{eqnarray}
We scale $2 {\rm Im} (\alpha^{12})$ and $(\omega/2\pi) \rho_{V, {\rm lin}}$ by a factor of 300 for a better layout together with ${\rm Re} (\alpha^{11} - \alpha^{22})$ and $(\omega/2\pi) \rho_{Q, {\rm lin}}$.  Note that $\rho_{V, {\rm lin}}$ is a good approximation for Faraday rotation coefficient at any temperature, while $\rho_{Q, {\rm lin}}$ is a good approximation for Faraday conversion coefficient only at low temperatures, not at high temperatures. The peak of Faraday conversion for monoenergetic particles leads to a similar peak at about $10^{11}$K for thermal distribution. Faraday conversion is much lower than the linear approximation predicts, if the temperature rises. We will show in the last section that the linear approximations lead to wrong predictions of circular polarization from Sgr A*, the difference being a factor of several.

Black long-dashed and dashed lines on the right panel show $(\omega/2\pi)\eta'_Q$ and $(\omega/2\pi)\eta'_V$, for $n_e=1$, respectively. The absorption coefficients $\eta'_Q$ and $\eta'_V$ are calculated according to Eq.~(\ref{sazonovab}) by substituting the thermal distribution function. Similar to the case of monoenergetic particles, our thermal calculations match the approximations well at high temperatures, while they become inaccurate in low temperatures. In practice one can just adopt the simple traditional approximations for synchrotron absorption coefficients.

\begin{figure}
\begin{center}
\includegraphics[scale = 1.1]{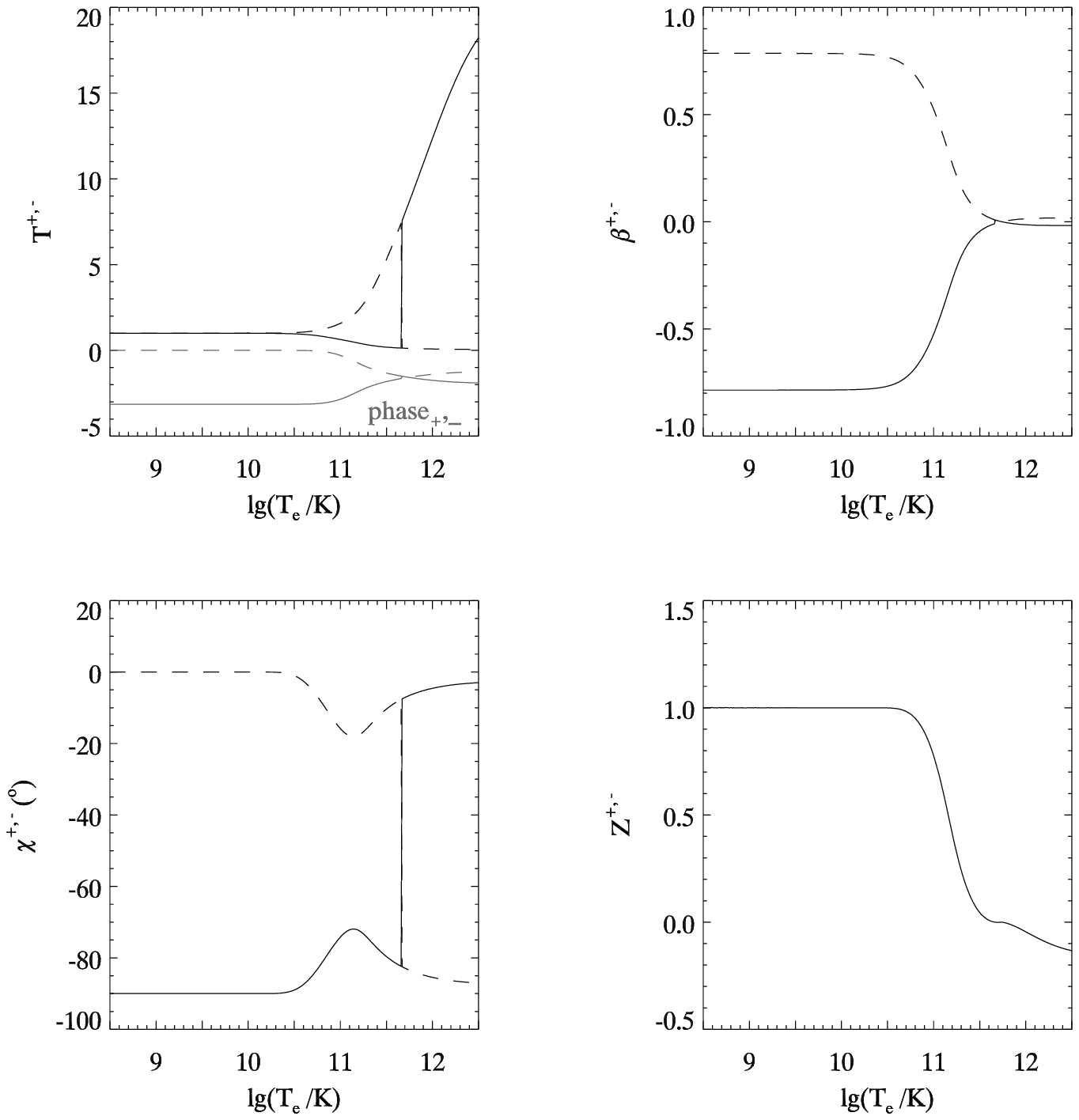}
\end{center}
\caption{Properties of two natural modes in plasma with thermal energy distribution.. Amplitudes $T^\pm$ (black) and phases $\varphi^\pm$  (grey) of complex axial ratio, ellipticity $\beta^\pm$,  EVPA $\chi^\pm$, and $Z^+=Z^-$ (eq.~\ref{Zmonster}) are shown on four panels, respectively.
}\label{figmodethermal}
\end{figure}

We show $T^\pm$ and phases, $\beta^\pm$,  $\chi^\pm$, and $Z^\pm$ on Figure~\ref{figmodethermal}.  In general, they are similar to those for $\delta$-function distribution discussed in \S~\ref{propnatmode}. Note that EVPA ($\chi$) deviates by as much as $\sim 20^\circ$ from $0^\circ$ (or $-90^\circ$) at $T_e \approx 10^{11}$K.

\subsubsection{Power-law distribution}
Number density per unit Lorentz factor is
\begin{eqnarray}
	N(\gamma) &=& \frac{b-1}{\gamma_{\rm min}^{1-b} - \gamma_{\rm max}^{1-b}}  \cdot n_e \gamma^{-b},  \quad \gamma_{\rm min} < \gamma < \gamma_{\rm max},  b > 1 
\end{eqnarray}
for power-law particle distribution, where $b$ is the energy-spectral index, set greater than $1$ as examples. Thus the distribution function is
\begin{eqnarray}
	f(\gamma) = \frac{N(\gamma)}{4 \pi \gamma^2 {\rm |v|}}   \quad=\quad  \frac{n_e}{4 \pi }  \cdot \frac{b-1}{\gamma_{\rm min}^{1-b} - \gamma_{\rm max}^{1-b}}  \cdot \frac{\gamma^{- (b + 1)}}{\sqrt{\gamma^2 - 1}}, 	\quad \gamma_{\rm min} < \gamma < \gamma_{\rm max}.
\end{eqnarray}
We set $\gamma_{\rm max}=+\infty$ in computations below for $b>1$. Then $f(\gamma_{\rm max} \to +\infty) \to 0$, and the response tensor becomes
\begin{eqnarray}
	\alpha^{ij}(k) = \frac{ n_e (b-1)}{4 \pi \gamma_{\rm min}^{1-b}} \left[ \int_{\gamma_{\rm min}}^{+\infty}  \frac{\gamma^{-b}}{\sqrt{\gamma^2 - 1}} \cdot \alpha^{ij}(k, \gamma) {\rm d} \gamma  +\frac{\gamma_{\rm min}^{-b}}{\sqrt{\gamma_{\rm min}^2 - 1}} \cdot \alpha^{ij}_{\rm B}(k, \gamma_{\rm min})  \right].
\end{eqnarray}

\begin{figure}
\begin{center}
\includegraphics[scale = 1.1]{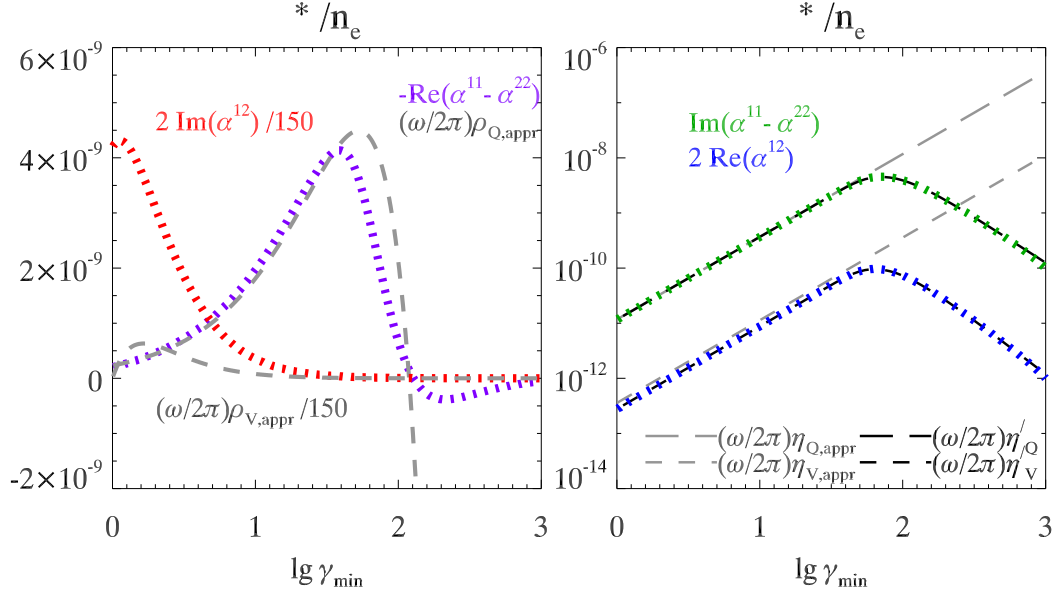}
\end{center}
\caption{Propagation and absorption coefficients for power-law energy distribution with index $b=2.5$.  {\it Left}: Faraday rotation and conversion. Linear approximations by \citet{melrose97b} are shown in grey dashed lines.   {\it Right}: Absorption coefficients. Traditional synchrotron approximations are shown in black solid lines, whereas the approximations by \citet{sazonov} are shown in grey dashed lines.}\label{figpl}
\end{figure}

We show the results of numerical integration for $b=2.5$ on Figure~\ref{figpl}. On the left panel, we also show approximations for propagation coefficients given by \citet{sazonov} for $n_e=1$ in dashed grey lines. The related formulae are
\begin{eqnarray}
	\rho_{Q, {\rm appr}} &=& 8.5\times 10^{-3} \cdot \frac{2}{(b-2)}\left[ \left( \frac{\omega}{\Omega_0 \sin\theta \gamma_{\rm min}^2} \right)^{(b-2)/2} - 1 \right]
\frac{(b-1)}{\gamma_{\rm min}^{1-b}} \cdot  \left( \frac{\Omega_0}{\omega}\sin\theta \right)^{(b+2)/2} \cdot \frac{2\pi}{\omega},   \nonumber\\
	\rho_{V, {\rm appr}} &=& 1.7\times 10^{-2}  \cdot \frac{\ln \gamma_{\rm min}}{(b+1) \gamma_{\rm min}^{b+1}}  \cdot \frac{(b-1)}{\gamma_{\rm min}^{1-b}} \cdot \frac{\Omega_0}{\omega} \cos\theta \cdot \frac{2\pi}{\omega}.
\end{eqnarray}
(Similar formulae can be also found in \citet{jones}.)

We scale $2 {\rm Im} (\alpha^{12})$ and $(\omega/2\pi) \rho_{V, {\rm appr}}$ by a factor of $150$ for a better layout.  Here $\rho_{V, {\rm appr}}$ is a good approximation of Faraday rotation coefficient only for large $\gamma_{\rm min}$. It significantly underestimates Faraday rotation, if the cut-off Lorentz factor is low.  The approximation $\rho_{Q, {\rm appr}}$ works well for Faraday conversion coefficient, if $\gamma_{\rm min} < 100$. It accurately describes the exact behavior including the peak.

On the right panel of Figure~\ref{figpl} long-dashed and dashed black lines show $(\omega/2\pi)\eta'_Q$ and $(\omega/2\pi)\eta'_V$ respectively  for $n_e=1$. The approximate absorption coefficients $\eta'_Q$ and $\eta'_V$ are computed based on Eq.~\ref{sazonovab} by substituting the power-law distribution function. In this case our calculation of ${\rm Im} (\alpha^{11} - \alpha^{22})$ and $2 {\rm Re} (\alpha^{12})$ match their traditional approximations well for all $\gamma_{\rm min}$. This is because the particles with high $\gamma$ play a large role in power-law distribution compared to the thermal, so that the inaccuracy at low $\gamma$'s is concealed.
Similar to the case of thermal distribution, one can adopt simple traditional approximations for synchrotron absorption.  We also show approximations for absorption coefficients given by \citet{sazonov} for $n_e=1$, in long-dashed and dashed grey lines. The corresponding expressions are
\begin{eqnarray}
	\eta_{Q,{\rm appr}} &=& 3.1 \times 10^{-4} \cdot \left( b + 2 \right) \Gamma\left( \frac{3 b + 2}{12} \right) \Gamma\left( \frac{3 b + 10}{12} \right)
\frac{(b-1)}{\gamma_{\rm min}^{1-b}} \cdot \left( 3 \frac{\Omega_0}{\omega} \sin\theta \right)^{(b+2)/2} \cdot \frac{2\pi}{\omega},  \nonumber \\
	\eta_{V,{\rm appr}} &=& 4.1 \times 10^{-4} \cdot \frac{b+3}{b+1} \Gamma\left( \frac{3 b + 7}{12} \right) \Gamma\left( \frac{3 b + 11}{12} \right)
\frac{(b-1)}{\gamma_{\rm min}^{1-b}} \cdot (b+2)\cot\theta \left( 3 \frac{\Omega_0}{\omega} \sin\theta \right)^{(b+3)/2} \cdot \frac{2\pi}{\omega}.
\end{eqnarray}
These approximations are good for $\gamma_{\rm min} < 100$, while they overestimate the absorption at larger $\gamma_{\rm min}$.

\begin{figure}
\begin{center}
\includegraphics[scale = 1.1]{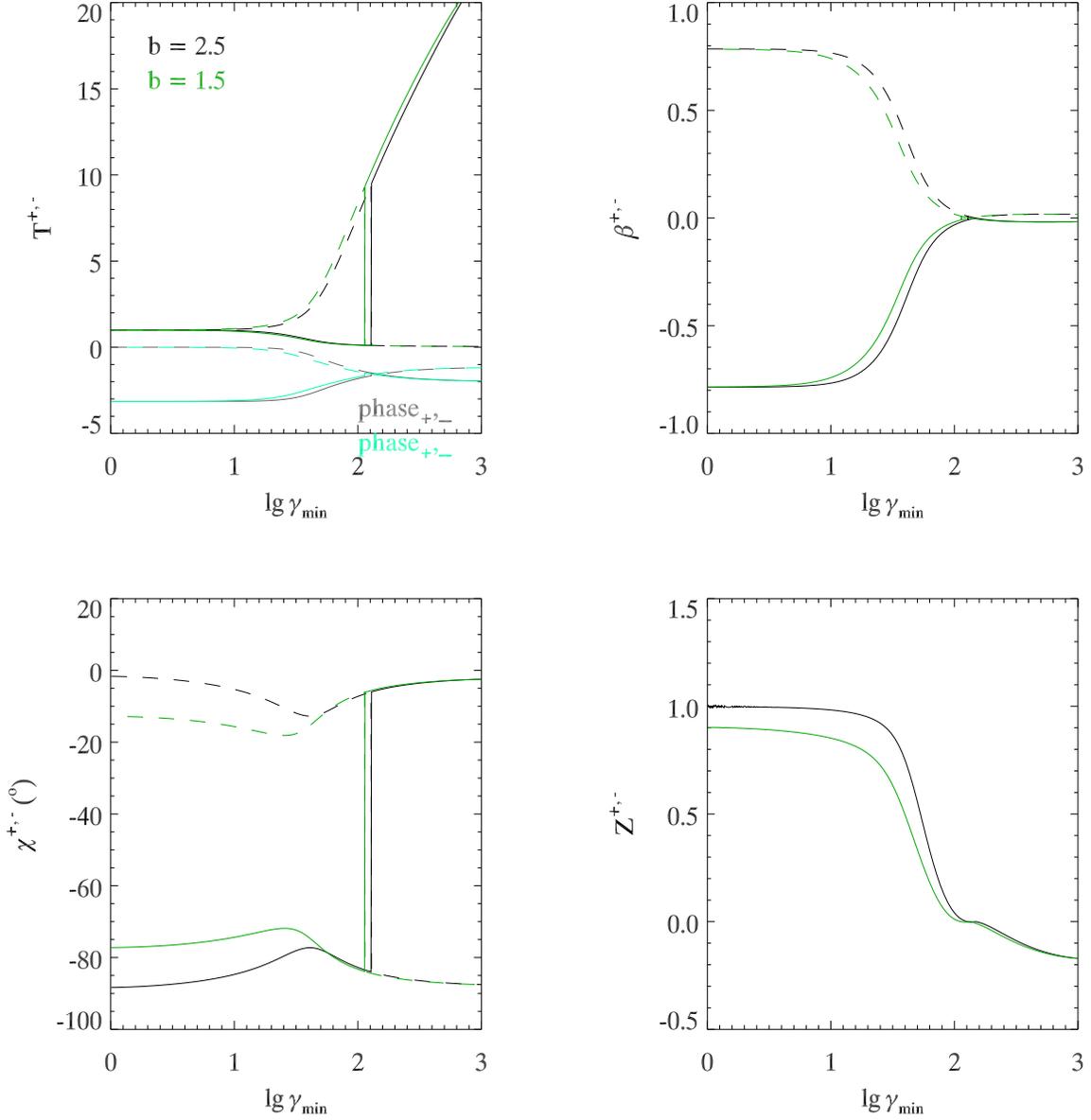}
\end{center}
\caption{Properties of two natural modes for the power-law particle distribution. Amplitudes $T^\pm$ (black) and phases $\varphi^\pm$  (grey) of complex axial ratio, ellipticity $\beta^\pm$,  EVPA $\chi^\pm$, and $Z^+=Z^-$ (eq.~\ref{Zmonster}) are shown in four panels, respectively, for $b=2.5$. Lines for $(^+)$-mode are solid and for $(^-)$-mode are dashed. The same quantities for the power-law distribution with $b=1.5$ are shown in green and cyan.}\label{figmodepl}
\end{figure}

We show $T^\pm$ and phases, $\beta^\pm$,  $\chi^\pm$, and $Z^\pm$ in black and grey on Figure~\ref{figmodepl} for power-law particle distribution with $b=2.5$.   The same quantities for power-law distribution with $b=1.5$ are shown in green and cyan for comparison. The plots are similar to those for monoenergetic distribution or thermal distribution. The curves for $b=1.5$ and $b=2.5$ almost coincide for high $\gamma_{\rm min}$. At low $\gamma_{\rm min}$, however, $\chi^\pm$ deviate more from $-90^\circ$ (or $0^\circ$) at $b=1.5$ compared to the case with $b=2.5$, and $Z^\pm \ne 1$ for $b=1.5$. This is because the particles with high Lorentz factors affect the results more for a lower spectral index.

\subsection{New approximate formulae for Faraday conversion/rotation coefficients}\label{simp}
We have shown that the traditional synchrotron approximations to the absorption coefficients are accurate and practical. On the contrary, simple linear approximations of the propagation coefficients have large errors at high electron energies. Although \citet{sazonov} also provided the integral expressions for the propagation coefficients in their Eq.~(2.3), \citet{jones} in their Eq.~(C16) as well, the Faraday rotation ($f^{(r)}$ therein) and especially Faraday conversion ($h^{(r)}$ therein) are inaccurate for high energies of electrons.

We devise a new set of approximate formulae for the complex response tensor in a plasma with monoenergetic particle distribution. The goal is to provide simple relations for accurate evaluation of propagation coefficients. We provide the numerical code in Mathematica 8 for the full evaluation at \url{http://astroman.org/Faraday_conversion/}, but we encourage the readers to use the simplified formulae for practical applications. These simplified formulae for Faraday rotation and Faraday conversion are computed for plasma with $\delta$-function energy distribution. They can be easily integrated over the arbitrary energy distribution. Similar approximate formulae for thermal particle distributions were computed in \citet{shcher_farad}.

\subsubsection{Computations for monoenergetic particle distribution}
It is non-trivial that good approximations to Faraday rotation/conversion coefficients exist in a three-dimensional parameter space of $\Omega_0/\omega$, $\theta$, and $\gamma$. However, we manage to find the formulae accurate to within $10\%$ at most reasonable combinations of parameters.

We define an auxiliary quantity
\begin{eqnarray}
	X_{\rm A} = \sqrt{ \frac{\sqrt{2} \sin \theta}{10^{-4}} \frac{\Omega_0}{\omega} }
\end{eqnarray}
and introduce four new expressions: $H_{\rm X}$, $H_{\rm B}$, $g_{\rm X}$, and $g_{\rm B}$ to approximate $-{\rm Re}(\alpha^{11}-\alpha^{22})$, $-{\rm Re}(\alpha_{\rm B}^{11}-\alpha_{\rm B}^{22})$, $2 {\rm Im}(\alpha^{12})$, and $2 {\rm Im}(\alpha_{\rm B}^{12})$, respectively.  The formulae are
\begin{eqnarray}
	H_{\rm X} (\gamma_0) &=&
	\left\{ \begin{array}{cc}
	 9.29 \cdot 10^{-9} \cdot \sqrt{1 - \gamma_0^{-1}}  \cdot  (X_{\rm A} \gamma_0)^{3.036},  &  X_{\rm A} \gamma_0 <  40,   \\
	 \\
	 \\
	- 0.000203 (X_{\rm A}\gamma_0)^{0.4343}	- 0.0013  \cos\left[  0.5646 \ln (X_{\rm A}\gamma_0) -  4.03 \right] \\
	+ 0.002 \exp\left[ - \frac{(\ln X_{\rm A}\gamma_0 - 4.2137)^2}{0.5429} \right] + 0.00083 \exp\left[ - \frac{(\ln X_{\rm A}\gamma_0 - 4.2137]^2}{0.2121} \right], &   X_{\rm A} \gamma_0  \ge   40,
\end{array} \right.    \nonumber \\
	&\quad& \nonumber \\
	H_{\rm B} (\gamma_0) &=&
	\left\{ \begin{array}{cc}
	4.67 \cdot10^{-9} \cdot (1 - \gamma_0^{-1})^{\frac{3}{2}} \cdot (X_{\rm A} \gamma_0)^{3.84},  &   X_{\rm A} \gamma_0 <  40, \\
	 \\
	 \\
	 0.864 - 0.2082 \left( \ln X_{\rm A} \gamma_0 \right)^2 + 0.0175 \left( \ln X_{\rm A} \gamma_0 \right)^4 - 0.000626 \left( \ln X_{\rm A} \gamma_0 \right)^6  \\
	 + 1.0175 \cdot 10^{-5} \left( \ln X_{\rm A} \gamma_0 \right)^8  - 7.686\cdot 10^{-8} \left( \ln X_{\rm A} \gamma_0 \right)^{10}   \\
	 - 0.01 \exp\left[ - \frac{ (\ln X_{\rm A}\gamma_0 - 4.0755)^2 }{0.0763} \right], &   X_{\rm A} \gamma_0  \ge  40,
\end{array} \right.    \nonumber \\
	&\quad& \nonumber \\
	g_{\rm X} (\gamma_0) &=&  1 - 0.4 \exp\left[ - \frac{(\ln X_{\rm A}\gamma_0 - 9.21)^2}{11.93} \right]  -  0.05 \exp\left[ - \frac{(\ln X_{\rm A}\gamma_0 - 5.76)^2}{1.33} \right]    +  0.075 \exp\left[ - \frac{(\ln X_{\rm A}\gamma_0 - 4.03)^2}{0.65} \right], \nonumber \\
	&\quad& \nonumber \\
	g_{\rm B} (\gamma_0) &=&  1 - 0.0045 (X_{\rm A}\gamma_0)^{0.52}.
\end{eqnarray}
The Faraday conversion/rotation coefficients for monoenergetic isotropic particle distribution with Lorentz factor $\gamma_0$ are calculated as
\begin{eqnarray}
	\rho_Q (\gamma_0) &=& - \frac{2\pi}{\omega} {\rm Re} (\alpha^{11} - \alpha^{22}) \quad= \frac{8\pi^2 e^2 }{m_e c\omega } \cdot X_{\rm A} \cdot  H_{\rm X} (\gamma_0) ,   \nonumber \\
	\rho_{Q, \rm B} (\gamma_0) &=& - \frac{2\pi}{\omega} {\rm Re} (\alpha_{\rm B}^{11} - \alpha_{\rm B}^{22})  \quad=  \frac{8\pi^2 e^2 }{m_e c\omega }  \cdot  H_{\rm B} (\gamma_0),    \nonumber \\
	\rho_V(\gamma_0) &=& 2 \cdot \frac{2\pi}{\omega} {\rm Im} (\alpha^{12} )  \quad=  \frac{8\pi^2 e^2 \Omega_0 \cos\theta}{m_e c\omega^2}  \cdot \ln \left(  \frac{1 + p_0/\gamma_0}{1 - p_0/\gamma_0} \right)  \cdot g_{\rm X} (\gamma_0),    \nonumber \\
	\rho_{V, \rm B} (\gamma_0) &=& 2 \cdot \frac{2\pi}{\omega} {\rm Im} (\alpha_{\rm B}^{12} )   \quad=  \frac{8\pi^2 e^2 \Omega_0 \cos\theta}{m_e c\omega^2}  \cdot   \left[   \gamma_0 \cdot \ln \left(  \frac{1 + p_0/\gamma_0}{1 - p_0/\gamma_0} \right)   - 2 p_0    \right]  \cdot g_{\rm B} (\gamma_0).    
\end{eqnarray}
Those for arbitrary particle distribution $f(\gamma)$ are calculated as
\begin{eqnarray}
	\rho_Q &=& \int_{\gamma_{\rm min}}^{\gamma_{\rm max}}  f(\gamma) \cdot \rho_Q (\gamma) {\rm d}\gamma  - f(\gamma_{\rm max}) \cdot \rho_{Q, {\rm B}} (\gamma_{\rm max}) + f(\gamma_{\rm min}) \cdot \rho_{Q, {\rm B}} (\gamma_{\rm min}),   \nonumber \\
	\rho_V &=& \int_{\gamma_{\rm min}}^{\gamma_{\rm max}}  f(\gamma) \cdot \rho_V (\gamma) {\rm d}\gamma  - f(\gamma_{\rm max}) \cdot \rho_{V, {\rm B}} (\gamma_{\rm max}) + f(\gamma_{\rm min}) \cdot \rho_{V, {\rm B}} (\gamma_{\rm min}).
\end{eqnarray}

\begin{figure}
\begin{center}
\includegraphics[scale = 1.1]{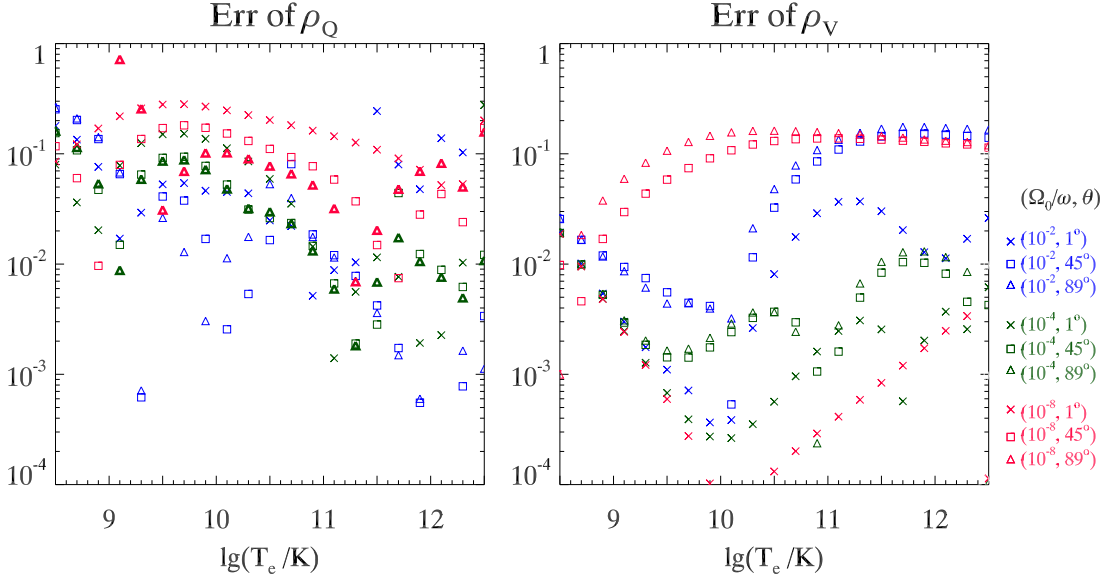}
\end{center}
\caption{ Accuracy checking for simplified formulae for Faraday conversion $\rho_{\rm Q}$ and Faraday rotation $\rho_{\rm V}$. See details in text. }\label{figerrthermal}
\end{figure}

The above formulae are accurate, with errors of several percent for $\delta$-distribution in general.  As a test, we choose $f(\gamma)$ in thermal distribution and calculate $(\omega/2\pi) \rho^T_{\rm Q}$ and $(\omega/2\pi) \rho^T_{\rm V}$ by them for a wide range of temperature. We then calculate $(\omega/2\pi) \rho^*_{\rm Q}$ and $(\omega/2\pi) \rho^*_{\rm V}$ by contour integrate formulae for thermal distribution, which is discussed in the Sec.4.2.2.  The errors of $\rho_{\rm Q}$ are defined as $|(\omega/2\pi) \rho^T_{\rm Q} - (\omega/2\pi) \rho^*_{\rm Q}|/|(\omega/2\pi) \rho^*_{\rm Q}|$, and those of $\rho_{\rm V}$, similarly, as $|(\omega/2\pi) \rho^T_{\rm V} - (\omega/2\pi) \rho^*_{\rm V}|/|(\omega/2\pi) \rho^*_{\rm V}|$. I.e., error of $1$ means the coefficient is accurate with a factor of 2.  We choose nine couples of parameters $(\frac{\Omega_0}{\omega}, \theta)$,  such as $(10^{-2}, 1^\circ)$, $(10^{-2}, 45^\circ)$, $(10^{-2}, 89^\circ)$, $(10^{-4}, 1^\circ)$, $(10^{-4}, 45^\circ)$, $(10^{-4}, 89^\circ)$, $(10^{-8}, 1^\circ)$, $(10^{-8}, 45^\circ)$, and $(10^{-8}, 89^\circ)$, as examples represented by different symbols (cross, square, and triangle) and different color (blue, green, and red), respectively.  As shown on Figure \ref{figerrthermal}, the errors are less than 1 (100\%) in general. For high temperatures at which synchrotron emission is effective ($> 10^{10}$K), the errors are as good as within 0.1 (10\%).

\subsubsection{Results for thermal distribution}
We have shown in \S~\ref{arbitthermal} that one can calculate the response tensor for monoenergetic particle first by integrating over the proper time (or phase), then integrate the result over the arbitrary particle distribution. For thermal distribution, one can change the integration order and analytically integrate over $\gamma$. Thus, the response tensor can be simplified to
\begin{eqnarray}
\label{respthermal}
	\alpha^{ij}(k) =  \frac{\imath n_e e^2}{ m_e c \Theta_e^{-2} K_2(\Theta_e^{-1})} \int^\infty_0 {\rm d} (\omega \xi)
\left( \dot{t}^{ij}(\xi) \frac{K_2({\mathcal R'}(\xi))}{{\mathcal R'}^2(\xi)} - \widetilde{T}^{ij}(\xi) \frac{K_3({\mathcal R'}(\xi))}{{\mathcal R'}^3(\xi)}  \right),
\end{eqnarray}
where
\begin{eqnarray}
	{\mathcal R'}(\xi) &=& \sqrt{  \Theta_e^{-2} - 2 \imath \Theta_e^{-1} \omega \xi  + \frac{\omega^2 \sin^2 \theta}{\Omega_0^2} [ 2 - 2\cos(\Omega_0 \xi) - \Omega_0^2 \xi^2 ]  }  \nonumber
\end{eqnarray}
and $\dot{t}^{ij}(\xi)$ and $\widetilde{T}^{ij}(\xi)$ are given in \S~\ref{permit}. See the corresponding derivation in Appendix C, which closely follows \citet{trubnikov}, \citet{melrose97a}, and \citet{swanson}. The integrals for thermal $\alpha^{ij}(k)$ in Eq.~(\ref{respthermal}) are easier to compute numerically. However, such simplification can only be done for thermal distribution. The detailed computations and discussion of Eq.~\ref{respthermal} can be found in \citet{shcher_farad}. Good fittings for thermal distribution, accurate within $10\%$ except with large $\Omega_0/\omega$ and large $\theta$, are also provided therein. We will provide their expressions for electrons in a present notation\footnote{Note, that the basis vectors in \citet{shcher_farad} are different, thus a different sign of $\rho_Q$.}:
\begin{eqnarray}
X_e &=& T_e\sqrt{\sqrt{2}\sin\theta \left(10^3\frac{\Omega_0}{\omega}\right)},   \nonumber \\
\rho_{V,th}  &=&  \frac{4\pi e^2\Omega_0}{m_e c\omega^2}\frac{K_0(T_e^{-1})}{K_2(T_e^{-1})}\cos\theta    \cdot  g(X_e),   \nonumber \\
\rho_{Q,th}  &=& \frac{2\pi e^2\Omega_0^2}{m_e c\omega^3}\left(\frac{K_{1}(T_e^{-1})}{K_2(T_e^{-1})}+6T_e\right)\sin^2\theta  \cdot h(X_e)
\end{eqnarray}
with approximate multipliers
\begin{eqnarray}\label{gaunt_diag}
g(X_e) &=& 1-0.11\ln(1+0.035X_e),  \nonumber \\
h(X_e) &=& 2.011\exp\left(-\frac{X_e^{1.035}}{4.7}\right)-\cos\left(\frac{X_e}2\right)\exp\left(-\frac{X_e^{1.2}}{2.73}\right)-0.011\exp\left(-\frac{X_e}{47.2}\right). 
\end{eqnarray}

\section{APPLICATIONS}\label{discus}

We computed the response tensor in uniformly magnetized relativistic plasmas with isotropic particle distributions. We found Faraday conversion, Faraday rotation, and absorption coefficients by numerical integration in the complex plane. Then we discussed the properties of natural modes of cyclo-synchrotron radiation and presented the results for specific electron energy distributions. We provided accurate practical fitting formulae for Faraday conversion and rotation. The method of complex plane integration allows to calculate both absorption and propagation coefficients consistently. In practice, formulae in \citet{sazonov} are good enough for absorption coefficients. Therefore, we focus on improving the calculations of Faraday conversion and rotation coefficients. Faraday conversion and rotation coefficients can be generally expressed as $\rho_Q \propto h(\gamma, \lambda B, \theta) n_e B^2 \lambda^3$ and $\rho_V \propto g(\gamma, \lambda B, \theta) n_e B \lambda^2$, where $h, g$ are functions of electron Lorentz factor $\gamma$, product of wavelength $\lambda$ and magnetic field $B=|{\bmath B}|,$ and angle $\theta$ between wavevector $\bmath k$ and magnetic field $\bmath B.$

Previous work \citep[e.g.,][]{melrose97b, qua_gru} has shown that $g \approx 1$, if $\gamma \approx 1$ for typical $\lambda,$ $B,$ $\theta,$ but $g \ll 1$ if $\gamma \gg 1$.  This means the Faraday rotation becomes less important, when the electrons become relativistic. Therefore, the direction of LP plane changes little. Our calculations suggest the function $g$ and the Faraday rotation coefficient were previously computed precisely, but function $h$ and Faraday conversion coefficient were not. The function $h$ grows from small to intermediate $\gamma$, but $h$ steeply decreases, if $\gamma$ grows more. This means the Faraday conversion also becomes less important when the electrons become relativistic, although the peak Lorentz factor depends on frequency ratio $\Omega_0/\omega$ and the particle distribution. In sum, as the absorption coefficients at a given frequency $\omega$ also decrease with $\gamma$, ultrarelativistic electrons interact less with radiation field.

We have shown that Faraday conversion coefficient should be computed more precisely. Let us now demonstrate that imprecise estimates of propagation effects result in large error in polarized simulated fluxes for accretion onto compact objects. We compare polarized spectra for two cases: accurate general relativistic polarized radiative transfer \citep{shcher_huang} and assuming that Faraday conversion and rotation coefficients are given by linear approximations. We test two types of dynamical models for Sgr A* accretion to prove the case. Both models assume thermal electron distribution.

\begin{figure}
\begin{center}
\includegraphics[scale = 0.5]{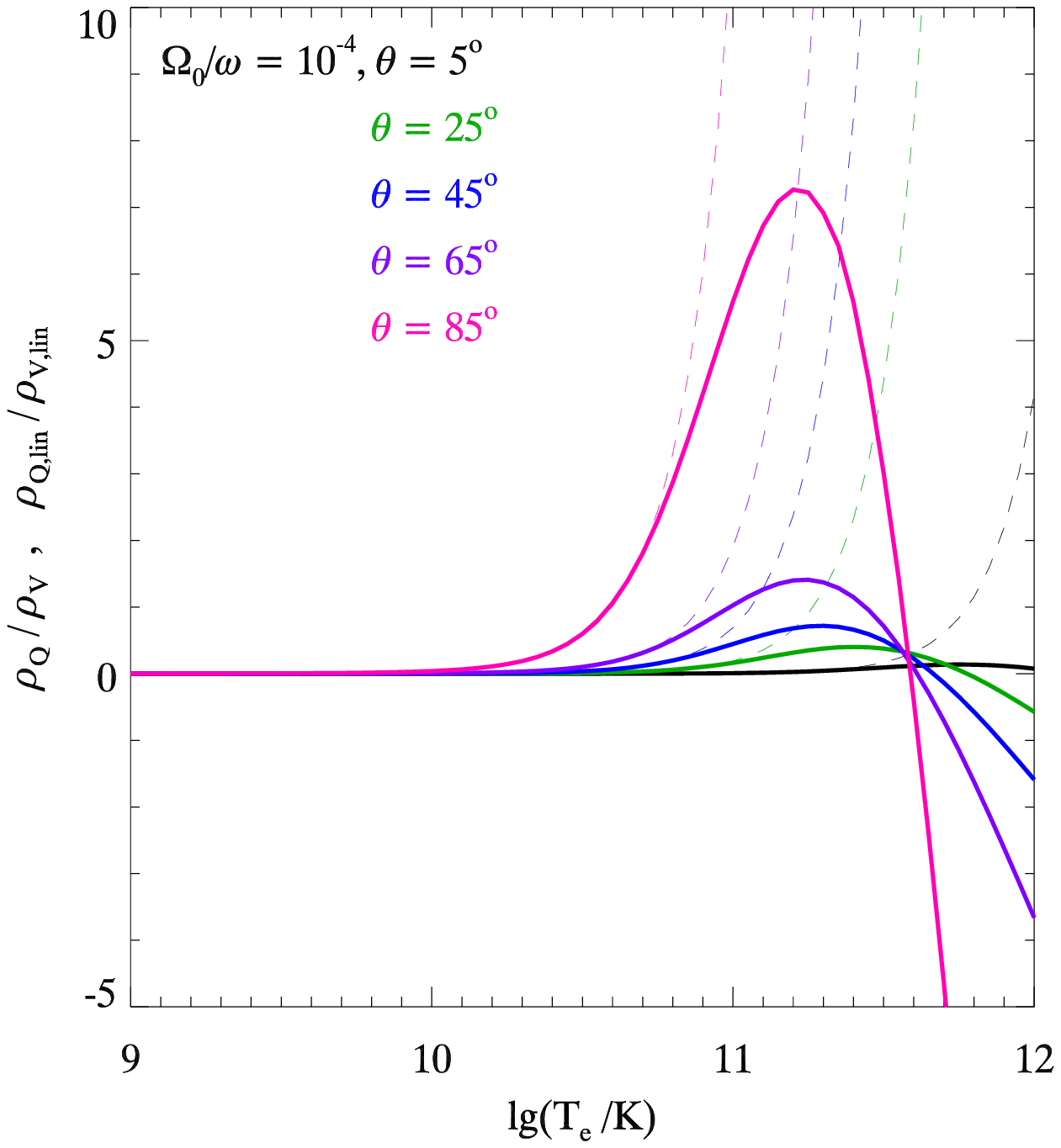}
\includegraphics[scale = 0.5]{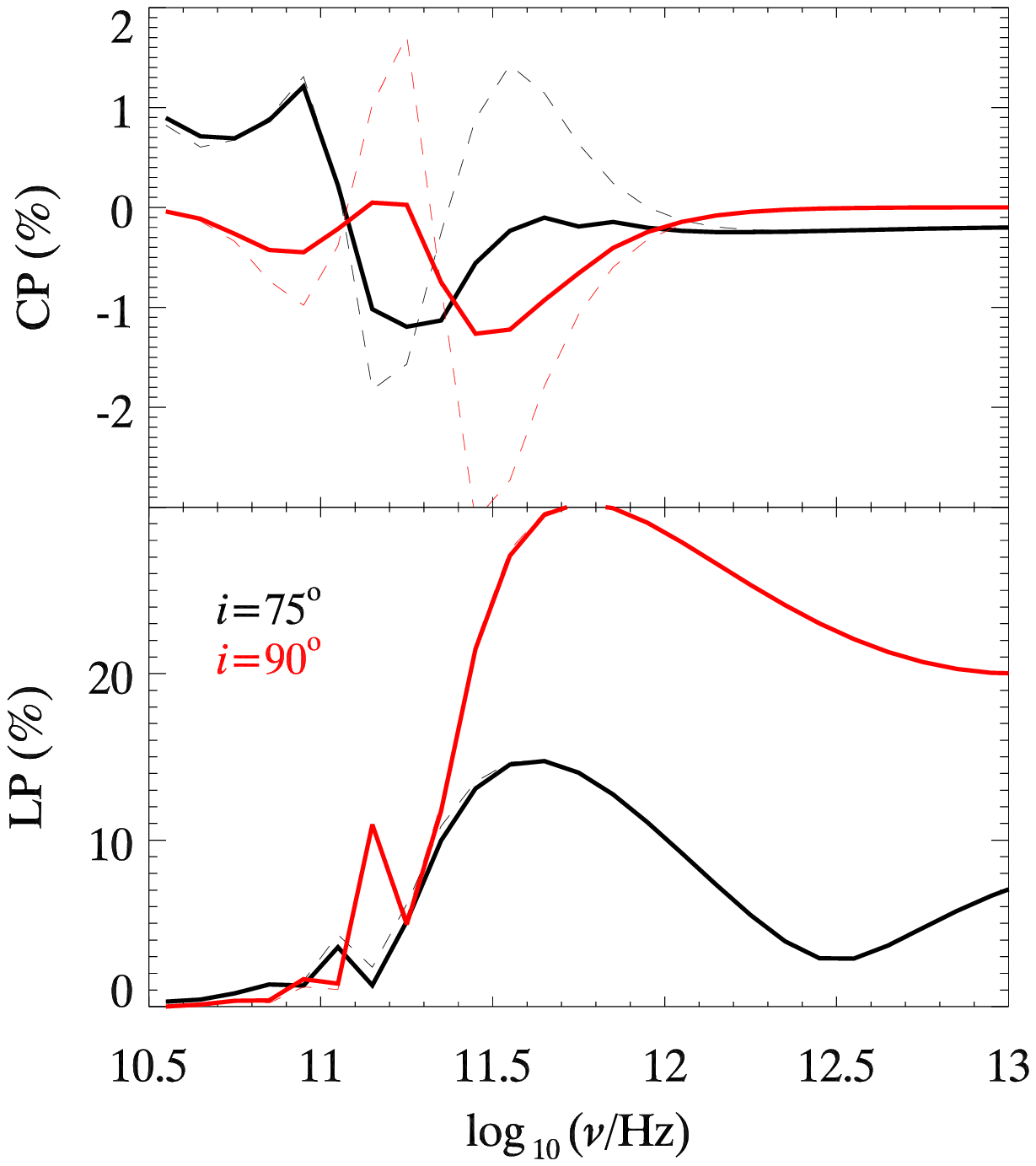}
\end{center}
\caption{{\it Left}: Examples of accurate ratios of $\rho_Q/\rho_V$ (solid) and their linear approximations $\rho_{Q, {\rm lin}} /\rho_{V, {\rm lin}}$ (dashed) for thermal electrons with $\Omega_0/\omega = 10^{-4}$ and different $\theta$.   {\it Right}: Simulated CP and LP fluxes based on accurate Faraday conversion/rotation (solid) and their linear approximations (dashed) for Sgr A* accretion model in \citep{huang09b} with inclination angle of $75^\circ$ (black) and $90^\circ$ (red).}\label{figsgracp}
\end{figure}

First, we adopt the analytic model from \citet{huang09b}.  This model is established for Sgr A* based on radiatively inefficient accretion flow (RIAF) solution with different temperatures in ions and electrons.  It reasonably fits Sgr A* polarized mm/sub-mm spectrum. On the left panel of Figure~\ref{figsgracp} we plot the accurate ratio $\rho_Q /\rho_V$ (solid lines) and the ratio $\rho_{Q, {\rm lin}} /\rho_{V, {\rm lin}}$ in linear approximation (dashed lines) as functions of electron temperature $T_e$ for $\Omega_0/\omega = 10^{-4}$ and different ${\bmath k} - {\bmath B}$ angles. On the right panel, we plot simulated circular and linear polarization spectra computed with accurate Faraday conversion and rotation coefficients (solid lines) and the results for the same dynamical model for assumed linear approximations of propagation coefficients (dashed lines). The inclination angles of the disk $i=75^\circ$ (black) and $i=90^\circ$ (red) are chosen.

\begin{figure}
\begin{center}
\includegraphics[scale = 0.70]{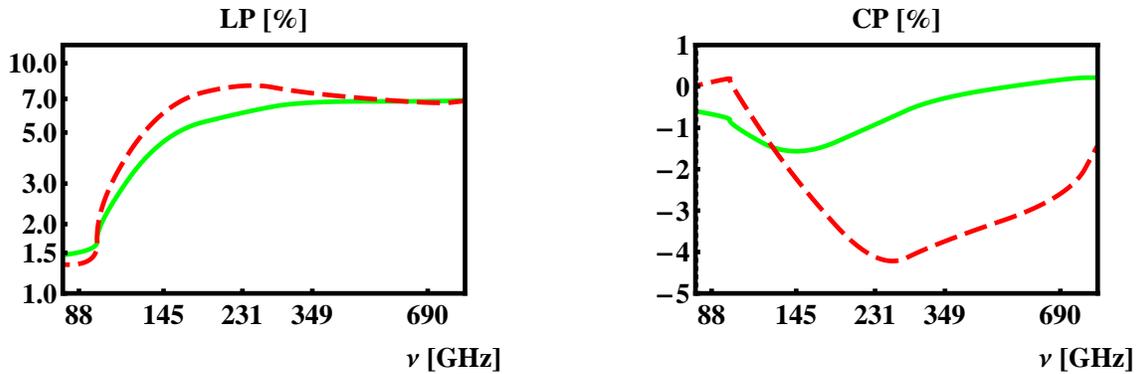}
\end{center}
\caption{Simulated LP (on the \textit{left}) and CP (on the \textit{right}) fractions for Sgr A* for various prescriptions of Faraday rotation and Faraday conversion: for linear approximations $\rho_{Q, {\rm lin}} /\rho_{V, {\rm lin}}$ (dashed red) and for accurate $\rho_V$ and $\rho_Q$ (solid green). We employ the best-fitting dynamical model from \citet{shcher_appl} with dimensionless spin $a_*=0.9.$}\label{fig_Roman}
\end{figure}

The frequency ratio $\Omega_0/\omega \sim 10^{-4}$ corresponds to sub-millimeter band close to the event horizon of Sgr A*. There the electron temperature is $T_e \sim 10^{10.5-11.5}$K. The Faraday conversion becomes strong ($\rho_Q /\rho_V > 1$) at large angles $\theta > 45^\circ$. One can also note from the right panel of Figure \ref{figthermal} that $\eta_V$ absorption coefficient is less than $1\%$ of $\eta_I$. Therefore emissivity in $V$ is equally weak according to Kirchhoff's Law. So Faraday conversion plays a major role in the production of circular polarization. However, $\rho_Q /\rho_V$ reaches a peak at a specific temperature around $10^{11}$K then decreases again, while $\rho_{Q, {\rm lin}} /\rho_{V, {\rm lin}}$ monotonically increases to much greater than $1$.  Therefore, the amplitudes of circular polarization predicted for  accurate $(\rho_Q,\rho_V)$ are less than half of those predicted for approximate $(\rho_{Q, {\rm lin}}, \rho_{V, {\rm lin}})$, although accurate and simplified propagation coefficients predict similar linear polarization.

Changes in simulated polarized fluxes are shown on Figure~\ref{fig_Roman} for the best-fitting Sgr A* accretion model from \citet{shcher_appl}.  This model is inspired by three-dimensional general relativistic magneto hydrodynamic simulations. The structure of magnetic field, velocity and density fields are taken directly from simulations. All free parameters are adjusted to achieve the best fit. Similarly to aforementioned analytic model, circular polarization in sub-millimeter band is significantly lower, when the precise Faraday conversion and rotation coefficients are adopted for the same dynamical model. Thus substantially lower predicted CP fractions, when the precise propagation coefficients are adopted, is a generic model-independent result. Our calculation showed that linear approximations of Faraday conversion/rotation are invalid for electrons with such high energy. They significantly overestimate the circular polarization for the relevant range of ratios $\Omega_0/\omega$. When fitting polarized observations and predicting polarized spectra one should adopt accurate Faraday conversion and rotation coefficients computed in \S~\ref{permit} or the simplified fitting formulae provided in \S~\ref{simp}.

The validity of linear approximations of Faraday conversion and rotation coefficients depends not only on the energy of electrons, but also on the frequency ratio $\Omega_0/\omega$. The observational frequency in near IR (NIR) is $\sim 10^3$ larger than that in sub-mm. The typical frequency ratio $\Omega_0/\omega$ is around $10^{-6} - 10^{-8}$, which yields the values of $(\rho_Q, \rho_V)$ similar to those of $(\rho_{Q, {\rm lin}}, \rho_{V, {\rm lin}})$ for $\gamma < 50$.  Thus, the linear approximations of Faraday conversion and rotation coefficients still adequately describe the correspondent effects in NIR.

Observations of sub-mm circular polarization from Sgr A* are one of today's big interests and challenges. CP fraction is only about $1\%$ in sub-mm, but was already detected at several frequencies \citep{munoz}. Still, there is substantial spread between different Sgr A* models \citep{shcher_appl} in CP fluxes at frequencies, where CP flux was not yet measured, for example $88$~GHz, $145$~GHz, and $690$~GHz. Observations at these frequencies can help to further discriminate between models of various types, which have different black hole spins.

\section*{ACKNOWLEDGMENTS}

The authors are grateful to Prof. Makoto Inoue, Dr. Rohta Takahashi, and Prof. Ramesh Narayan for useful suggestions.  \textbf{The comments of anonymous referee helped to improve presentation of findings.} This work was supported in part by the National Natural Science Foundation of China (grants 10625314, 10633010 and 10821302), the Knowledge Innovation Program of the Chinese Academy of Sciences (Grant No. KJCX2-YW-T03), the National Key Basic Research Development Program of China (No. 2007CB815405), and the CAS/SAFEA International Partnership Program for Creative Research Teams.  LH acknowledges the support by China Postdoctoral Science Foundation (grant 20090450822) and Regular Postdoctoral Research Fellowship in Academia Sinica.  RVS acknowledges the support by NASA grants NNX08AX04H and NNX11AE16G.

\label{lastpage}

\newpage

\appendix{\textbf{Appendix A:   Definitions and derivations}}\label{appendix_response}
\\

Here we summarize all definitions and derivations related to the response tensor. They are similar to those in \citep{melrose97a,melrose97b}, except for the signature of the metric tensor and the geometry of the coordinate system. We list all related tensors which are different from those in Melrose's work.

Metric tensor in Minkowski space-time: $g^{\mu\nu} = {\rm diag} [-1,1,1,1]$. \\
Wave vector: $k^\mu = \omega (1, 0, 0, 1)^{\rm T}$. \\
Tensor of magnetostatic field:
\begin{eqnarray}
	F_0^{\mu\nu} = B f^{\mu\nu},  \qquad B \quad=\quad \left( \frac{1}{2} F_0^{\mu\nu} F_{0 \mu\nu} \right)^{1/2}, \qquad
	f_{\mu\nu} = \left( \begin{array}{cccc}
		0 & 0 & 0 & 0 \\
		0 & 0 & \cos\theta & \sin\theta \\
		0 & -\cos\theta & 0 & 0 \\
		0 & - \sin\theta & 0 & 0 \end{array} \right).\nonumber
\end{eqnarray}
Auxiliary tensor for electron velocity perturbations:
\small
\begin{eqnarray}
	\dot{t}^{\mu\nu} (\tau)	=
	\left( \begin{array}{cccc}
		-1 & 0 & 0 & 0 \\
		0 & \cos\Omega_0 \tau & - \cos\theta \sin\Omega_0 \tau & - \sin\theta \sin\Omega_0 \tau  \\
		0 & \cos\theta \sin\Omega_0 \tau & \sin^2\theta +\cos^2\theta \cos\Omega_0 \tau& - \sin\theta \cos\theta + \sin\theta \cos\theta \cos\Omega_0 \tau \\
		0 & \sin\theta \sin\Omega_0 \tau & - \sin\theta \cos\theta + \sin\theta \cos\theta \cos\Omega_0 \tau & \cos^2\theta +\sin^2\theta \cos\Omega_0 \tau \end{array} \right).  \nonumber
\end{eqnarray}
\normalsize
Auxiliary tensor for electron position perturbations:
\small
\begin{eqnarray}
	T^{\mu\nu} (\tau) = t^{\mu\nu} (\tau) - t^{\mu\nu} (0) =
	\left( \begin{array}{cccc}
		- \tau & 0 & 0 & 0 \\
		0 & \frac{\sin\Omega_0 \tau}{\Omega_0} & - \cos\theta \frac{1 - \cos\Omega_0 \tau}{\Omega_0} & - \sin\theta \frac{1 - \cos\Omega_0 \tau}{\Omega_0} \\
		0 & \cos\theta \frac{1 - \cos\Omega_0 \tau}{\Omega_0} & \sin^2\theta \tau + \cos^2\theta \frac{\sin\Omega_0 \tau}{\Omega_0} & - \sin\theta \cos\theta \tau + \sin\theta \cos\theta \frac{\sin\Omega_0 \tau}{\Omega_0} \\
		0 & \sin\theta \frac{1 - \cos\Omega_0 \tau}{\Omega_0} & - \sin\theta \cos\theta \tau + \sin\theta \cos\theta \frac{\sin\Omega_0 \tau}{\Omega_0} & \cos^2\theta \tau +\sin^2\theta \frac{\sin\Omega_0 \tau}{\Omega_0} \end{array} \right). \nonumber
\end{eqnarray}
\normalsize
In 6-dimensional phase space: electron velocity ${\rm v}$, electron momentum $m_e {\rm p}$, and distribution function of electrons ${\rm f}({\rm p})$. \\
In 8-dimensional phase space: Lorentz factor $\gamma = 1/\sqrt{1 - {\rm |v|}^2}$, velocity of electron ${\mathcal U}^\mu = (\gamma, {\rm u}) = (\gamma, \gamma {\rm v})$, momentum ${\mathcal P}^\mu = (\gamma m_e, m_e {\rm p}) = (\gamma m_e, \gamma m_e {\rm v})$, and distribution function of electrons  ${\mathcal F}({\mathcal P}) = 2 m_e \delta \left({\mathcal P}^2 + m_e^2) {\rm f}({\rm p} \right)$.  \\
The general form of the response tensor for an arbitrary isotropic distribution of electrons is
\begin{eqnarray}
	\alpha^{\mu\nu}(k) = \imath e^2 \int {\rm d}^4 {\mathcal P}(\tau) \int^\infty_0 {\rm d} \xi \cdot {\mathcal U}^\mu(\tau) e^{\imath k[X(\tau-\xi) - X(\tau)]}
k{\mathcal U}(\tau-\xi) \left(g^{\lambda\nu} - \frac{k^\lambda {\mathcal U}^\nu (\tau-\xi)}{k{\mathcal U}(\tau-\xi)}\right) \dot{t}^\sigma_\lambda (\tau-\xi)  \frac{\partial {\mathcal F}({\mathcal P})}{\partial {\mathcal P}^\sigma (0)}.  \nonumber
\end{eqnarray}
\\

\appendix{\textbf{Appendix B:  Derivation of the response tensor for monoenergetic particle distribution}}\label{appendix_delta}
\\

Let us introduce $\widetilde{\mathcal U}^\mu = [1,0,0,0]$ --- 4-velocity in the rest frame of plasma. Then
\begin{eqnarray}
	\alpha^{\mu\nu}(k) &=& \imath e^2 \int {\rm d}^4 {\mathcal P}(\tau)  \nonumber \\
	&\times& \int^\infty_0 {\rm d} \xi {\mathcal U}^\mu(\tau) e^{\imath k[X(\tau-\xi) - X(\tau)]}
k{\mathcal U}(\tau-\xi) \left(g^{\lambda\nu} - \frac{k^\lambda {\mathcal U}^\nu (\tau-\xi)}{k{\mathcal U}(\tau-\xi)}\right) \dot{t}^\sigma_\lambda (\tau-\xi)
\left[ 2 m_e \delta \left({\mathcal P}^2 + m_e^2 \right) \right] m_e^{-3} \frac{\partial {\rm f}(\gamma)}{\partial {\mathcal P}^\sigma (0)}  \nonumber \\
	&=&  \imath e^2 \int {\rm d}^4 {\mathcal P}(\tau)   \nonumber \\
	&\times& \int^\infty_0 {\rm d} \xi {\mathcal U}^\mu(\tau) e^{\imath k[X(\tau-\xi) - X(\tau)]}
k{\mathcal U}(\tau-\xi) \left(g^{\lambda\nu} - \frac{k^\lambda {\mathcal U}^\nu (\tau-\xi)}{k{\mathcal U}(\tau-\xi)}\right)   \dot{t}^\sigma_\lambda (\tau-\xi) \widetilde{\mathcal U}_\sigma
\left[ 2 m_e \delta \left({\mathcal P}^2 + m_e^2 \right) \right] m_e^{-4} \frac{{\rm d} {\rm f}(\gamma)}{{\rm d} \gamma}  \nonumber \\
	&=&  \frac{\imath e^2}{m_e} \int {\rm d}^4 {\mathcal P}(\tau) \left[ 2 m_e \delta \left({\mathcal P}^2 + m_e^2 \right) \right] m_e^{-3} \frac{{\rm d} {\rm f}(\gamma)}{{\rm d} \gamma}
\int^\infty_0 {\rm d} \xi {\mathcal U}^\mu(\tau) e^{\imath k[X(\tau-\xi) - X(\tau)]} \left[ k{\mathcal U}(\tau-\xi) \widetilde{\mathcal U}^\nu  - k\widetilde{\mathcal U} {\mathcal U}^\nu(\tau-\xi) \right]     \nonumber \\
	&\quad& \nonumber \\
	&\quad& ( {\rm let} \quad \tau=0 )  \nonumber \\
	&=& - \frac{e^2}{m_e} \int {\rm d}^4 {\mathcal P}(\tau) \left[ 2 m_e \delta \left({\mathcal P}^2 + m_e^2 \right) \right] m_e^{-3} \frac{{\rm d} {\rm f}(\gamma)}{{\rm d} \gamma}
\left[ {\mathcal U}^\mu \widetilde{\mathcal U}^\nu + \imath k\widetilde{\mathcal U} \int^\infty_0 {\rm d} \xi {\mathcal U}^\mu \dot{t}^{\nu\sigma}(-\xi) {\mathcal U}_\sigma e^{\imath k_\lambda T^{\lambda\sigma}(-\xi) {\mathcal U}_\sigma} \right]  \nonumber \\
	&=&  - \frac{e^2}{m_e}  \int {\rm d}^3 {\rm p}  \left( \frac{1}{\gamma} \right) \frac{{\rm d} {\rm f}(\gamma)}{{\rm d} \gamma}
\left[ {\mathcal U}^\mu \widetilde{\mathcal U}^\nu + \imath k\widetilde{\mathcal U} \int^\infty_0 {\rm d} \xi {\mathcal U}^\mu \dot{t}^{\nu\sigma}(-\xi) {\mathcal U}_\sigma e^{- \imath k_\lambda T^{\sigma\lambda}(\xi)  {\mathcal U}_\sigma} \right] \nonumber
\end{eqnarray}
\begin{eqnarray}
	&=&    - \frac{e^2}{m_e c}   \int {\rm d}^3 {\rm p} \frac{{\rm d} f(\gamma)}{{\rm d} \gamma}  \widetilde{\mathcal U}^\mu \widetilde{\mathcal U}^\nu      +    \frac{\imath e^2 \omega}{m_e c}       \int^\infty_0 {\rm d} \xi \dot{t}^\nu_\sigma (-\xi) \frac{\partial^2}{\partial {\mathcal S}_\mu \partial {\mathcal S}_\sigma} \left[  \int {\rm d}^3 {\rm p}    \left(  - \frac{1}{\gamma}  \frac{{\rm d} f(\gamma)}{{\rm d} \gamma} {\rm e}^{- \imath {\mathcal R}(\xi) {\mathcal U} + {\mathcal S} {\mathcal U}}  \right) \right]_{{\mathcal S}_\mu=0},
\end{eqnarray}
which coincides with Eq.~\ref{resp0}. We denoted ${\mathcal R}^\mu(\xi) = k_\lambda T^{\mu\lambda}(\xi)$ and introduced an auxiliary variable ${\mathcal S}_\mu$.
\\

\appendix{\textbf{Appendix C: Derivation of the response tensor for thermal particle distribution}}\label{appendix_thermal}
\\

Substituting the thermal distribution into Eq.~\ref{respform} we obtain
\begin{eqnarray}
	I (\xi, s) = \frac{n_e}{\Theta_e K_2(\Theta_e^{-1})}   \int_1^{+\infty}  {\rm e}^{ - (\Theta_e^{-1} - \imath \omega \xi + s_0) \gamma}
\left[  ( \imath \omega \xi  + s_0 ) \cdot \frac{\sin \left( |{\rm\bf R} + \imath {\rm\bf s}| |{\rm\bf p}| \right)}{|{\rm\bf R} + \imath {\rm\bf s}|}
+ \frac{\gamma}{\sqrt{ \gamma^2 -1 }}  \cdot \cos \left( |{\rm\bf R} + \imath {\rm\bf s}| |{\rm\bf p}| \right)  \right] {\rm d}\gamma,
\end{eqnarray}
where $|{\rm\bf p}| = \sqrt{\gamma^2 -1}$.
We can simplify the above formula as follows:
\begin{eqnarray}
	\frac{\Theta_e K_2(\Theta_e^{-1})}{n_e} \cdot I (\xi, s) &=&  \left[  - \frac{\imath \omega \xi + s_0}{(\Theta_e^{-1} - \imath \omega \xi + s_0)}  -  1   \right]   \int_0^{+ \infty}    \frac{\sin( |{\rm\bf R} + \imath {\rm\bf s}| |{\rm\bf p}| )}{ |{\rm\bf R} + \imath {\rm\bf s}| }  \cdot  {\rm d} {\rm e}^{ - ( \Theta_e^{-1} - \imath \omega \xi  + s_0 ) \gamma} \nonumber \\
	&=& \frac{1}{(\Theta_e^{-1} - \imath \omega \xi + s_0)}  \int_0^{+ \infty}   {\rm e}^{ - ( \Theta_e^{-1} - \imath \omega \xi  + s_0 ) \gamma}  \cos ( |{\rm\bf R} + \imath {\rm\bf s} | |{\rm\bf p}| )  \cdot  {\rm d} |{\rm\bf p}|	  \nonumber \\
	&=& \frac{1}{2 (\Theta_e^{-1} - \imath \omega \xi + s_0)}  \int_{-\infty}^{+ \infty}   {\rm e}^{ - ( \Theta_e^{-1} - \imath \omega \xi  + s_0) \gamma} \cdot  {\rm e}^{ |{\rm\bf R} + \imath {\rm\bf s} | |{\rm\bf p}| }  \cdot  {\rm d} |{\rm p}|   \nonumber \\
	&\quad& \nonumber \\
	&\quad& ( {\rm let} \quad |{\rm\bf p}| = \sinh z, \quad \gamma = \cosh z  )   \nonumber \\
	&=&  \frac{1}{2 (\Theta_e^{-1} - \imath \omega \xi + s_0)}  \int_{-\infty}^{+ \infty}   {\rm e}^{ - ( \Theta_e^{-1} - \imath \omega \xi  + s_0 ) \cosh z + |{\rm\bf R} + \imath {\rm\bf s}| \sinh z}  \cosh z{\rm d} z   \nonumber \\
	&=&  \frac{\partial}{\partial (\Theta_e^{-1} - \imath \omega\xi + s_0)^2}  \int_{-\infty}^{+\infty}  {\rm d} z
\quad {\rm e}^{  - \sqrt{ ( \Theta_e^{-1} - \imath \omega \xi + s_0)^2 +  |{\rm\bf R}  + \imath {\rm\bf s}|^2 }  \cosh \left[ z - \imath \tan^{-1} \left( \frac{ |{\rm\bf R}  + \imath {\rm\bf s}| }{\Theta_e^{-1} - \imath \omega \xi + s_0} \right) \right]  }   \nonumber \\
	&=&  \frac{\partial}{\partial (\Theta_e^{-1} - \imath \omega\xi + s_0)^2} \left[  2 K_0\left( \sqrt{ ( \Theta_e^{-1} - \imath \omega \xi  + s_0 )^2 +  |{\rm\bf R}  + \imath {\rm\bf s}|^2 } \right)  \right]   \nonumber \\
	&\quad& \nonumber \\
	&\quad&	( \quad {\rm let} \quad {\mathcal R'}(\xi) = \sqrt{ ( \Theta_e^{-1} - \imath \omega \xi  + s_0 )^2 +  |{\rm\bf R}  + \imath {\rm\bf s}|^2 }  \quad)   \nonumber \\
	&=& \frac{K_1\left( {\mathcal R'}(\xi) \right)}{{\mathcal R'}(\xi)}.
\end{eqnarray}
Following Eq.~\ref{resp0} and using
\begin{eqnarray}
	\frac{1}{{\mathcal R'}(\xi)} \frac{\rm d}{{\rm d} {\mathcal R'}(\xi)} \left[ \frac{K_n({\mathcal R'}(\xi))}{{\mathcal R'}(\xi)^n} \right] &=& - \frac{K_{n+1}({\mathcal R'}(\xi))}{{\mathcal R'}(\xi)^{n+1}} \nonumber
\end{eqnarray}
we have
\begin{eqnarray}
	\alpha^{ij}(k) =  \frac{\imath n_e e^2}{ m_e c \Theta_e^{-2} K_2(\Theta_e^{-1})} \int^\infty_0 {\rm d} (\omega \xi)
\left( \dot{t}^{ij}(\xi) \frac{K_2({\mathcal R'}(\xi))}{{\mathcal R'}^2(\xi)} - \widetilde{T}^{ij}(\xi) \frac{K_3({\mathcal R'}(\xi))}{{\mathcal R'}^3(\xi)}  \right),
\end{eqnarray}
which is Eq.~\ref{respthermal}.
\end{document}